\PassOptionsToPackage{unicode}{hyperref}
\PassOptionsToPackage{bookmarks=false}{hyperref}
\documentclass[preprint2,numberedappendix]{emulateapj-rtx4}

\usepackage{etex}
\usepackage{graphicx} %
\graphicspath{{./fig/}{./png/}}
\usepackage{bm} %
\usepackage[dvipsnames]{xcolor}
\usepackage[draft,textsize=tiny,color=ForestGreen!30]{todonotes}
\usepackage{paralist}
\usepackage{amssymb,amsfonts,amsmath}%
\usepackage{array,booktabs,units}
\usepackage{bbold} 
\usepackage{tikz}

\renewcommand{\max}{{\rm max}}
\newcommand{\btimes}{{\bm\times}}
\newcommand{\alphaem}{\ensuremath{\alpha_{\rm em}}}
\newcommand{\tauc}{\ensuremath{\tilde{t}}} 
\newcommand{\helicity}{\ensuremath{\chi_{\rm m}}}
\newcommand{\pfrac}[2]{\frac{\partial #1}{\partial #2}}%
\newcommand{\tot}{{\text{tot}}}
\newcommand{\el}{{\text{el}}}
\newcommand{\mat}{{\text{mat}}}
\newcommand{\ohm}{{\text{Ohm}}}
\newcommand{\bec}[1]{\mbox{\boldmath $ #1$}}
\newcommand{\nab}{\bm{\nabla}}
\newcommand{\AAA}{\bm{A}}
\newcommand{\BB}{\bm{B}}

\newcommand{\EE}{\bm{E}}
\newcommand{\JJ}{\bm{J}}
\newcommand{\UU}{\bm{U}}
\newcommand{\VV}{\bm{V}_0}
\newcommand{\bb}{\bm{b}}
\newcommand{\uu}{\bm{u}}
\newcommand{\ff}{\bm{f}}
\newcommand{\SSSS}{\mbox{\boldmath ${\sf S}$} {}}
\newcommand{\meanrho}{\overline{\rho}}
\newcommand{\meanPhi}{\overline{\Phi}}
{}
{}
{}

{}
{}
\newcommand{\meanEMF}{\overline{\mbox{\boldmath ${\cal E}$}}{}}{}
{}
{}
{}
{}
{}
\newcommand{\meanAA}{\overline{\mbox{\boldmath $A$}}{}}{}
\newcommand{\meanBB}{\overline{\mbox{\boldmath $B$}}{}}{}
\newcommand{\meanEE}{\overline{\mbox{\boldmath $E$}}{}}{}
{}
{}
{}
{}
{}
{}
{}
\newcommand{\meanUU}{\overline{\bm{U}}}

{}
{}
{}
\newcommand{\meanA}{\overline{A}}
\newcommand{\meanB}{\overline{B}}

\newcommand{\meanS}{\overline{S}}

\newcommand{\meanv}{\overline{v}}

\newcommand{\meanmu}{\overline{\mu}}

\def\oA{\omega_{\rm A}}
\def\geta{\gamma_\eta}
\def\gmu{\gamma_\mu}

\def\Pm{{\rm Pr}_{_{M}}}
\def\Rm{{\rm Re}_{_{M}}}
\def\half{{\textstyle{1\over2}}}
\newcommand{\EQA}{\begin{eqnarray}}
\newcommand{\ENA}{\end{eqnarray}}

\usepackage{hyperref}
\newcommand\myshade{85}
\colorlet{mylinkcolor}{violet}
\colorlet{mycitecolor}{Aquamarine}
\colorlet{myurlcolor}{YellowOrange}

\hypersetup{
 linkcolor  = mylinkcolor,
 citecolor  = mycitecolor!\myshade!black,
 urlcolor   = myurlcolor!\myshade!black,
colorlinks = true,
}
\begin{document}
\title{Laminar and turbulent dynamos in chiral magnetohydrodynamics-I:
  Theory}
\author{Igor~Rogachevskii$^{1,2,3}$}
\email{gary@bgu.ac.il}

\author{Oleg~Ruchayskiy$^{4}$}

\author{Alexey~Boyarsky$^{5}$}

\author{J\"{u}rg~Fr\"{o}hlich$^{6}$}

\author{Nathan~Kleeorin$^{1,2}$}

\author{Axel~Brandenburg$^{2,3,7,8}$}

\author{Jennifer~Schober$^{2}$}

\medskip
\medskip
\affiliation{
$^1$Department of Mechanical Engineering,
 Ben-Gurion University of the Negev, P.O. Box 653, Beer-Sheva
 84105, Israel \\
$^2$Nordita, KTH Royal Institute of Technology
 and Stockholm University, Roslagstullsbacken 23,
 SE-10691 Stockholm, Sweden \\
$^3$Laboratory for Atmospheric and Space Physics,
   University of Colorado, 3665 Discovery Drive, Boulder, CO 80303, USA \\
$^4$Discovery Center, Niels Bohr Institute, Blegdamsvej 17,
 DK-2100 Copenhagen, Denmark \\
$^5$Instituut-Lorentz for Theoretical Physics, Universiteit Leiden,
 Niels Bohrweg 2, 2333 CA Leiden, The Netherlands \\
$^6$Institute of Theoretical Physics, ETH H\"{o}nggerberg,
 CH-8093 Zurich, Switzerland \\
$^7$Department of Astronomy, AlbaNova University Center,
    Stockholm University, SE-10691 Stockholm, Sweden\\
$^8$JILA and Department of Astrophysical and Planetary Sciences,
    Box 440, University of Colorado, Boulder, CO 80303, USA
}

\submitted{Astrophys. J. 846, 153 (2017)}
\date{Received 2017 April 30; revised 2017 August 19; accepted 2017 August 22; published 2017 September 13}

\begin{abstract}
The magnetohydrodynamic (MHD) description of plasmas with relativistic
particles necessarily includes an additional new field, the chiral
chemical potential associated with the axial charge (i.e., the
number difference between right- and left-handed relativistic fermions).
This chiral chemical potential gives rise to a contribution to the electric current density of the plasma (\emph{chiral magnetic effect}).
We present a self-consistent treatment of the \emph{chiral
MHD equations}, which include the back-reaction of the magnetic
field on a chiral chemical potential and its interaction
with the plasma velocity field.
A number of novel phenomena are exhibited.
First, we show that the chiral magnetic effect decreases the frequency of the Alfv\'{e}n wave for incompressible flows, increases the frequencies of the Alfv\'{e}n wave and of the fast magnetosonic wave for compressible flows, and decreases the frequency of the slow magnetosonic wave.
Second, we show that, in addition to the well-known laminar
chiral dynamo effect, which is not related to fluid motions,
there is a dynamo caused by the joint action of velocity shear
and chiral magnetic effect.
In the presence of turbulence with vanishing mean kinetic helicity,
the derived mean-field chiral MHD equations
describe turbulent large-scale dynamos caused by the chiral alpha effect,
which is dominant for large fluid and magnetic Reynolds numbers.
The chiral alpha effect is due to an interaction of
the chiral magnetic effect and fluctuations of the small-scale
current produced by tangling magnetic fluctuations
(which are generated by tangling of the large-scale magnetic field
by sheared velocity fluctuations).
These dynamo effects may have interesting consequences in the dynamics of the early
universe, neutron stars, and the quark--gluon plasma.
\end{abstract}


\maketitle

\section{Introduction}
\label{sec:introduction}

Hydrodynamics is a universal description of a vast variety of systems ranging from astrophysical environments to biological systems.
A hydrodynamic description is possible whenever a form of \emph{local thermal equilibrium} prevails in macroscopically large regions of space.
Variables describing the thermodynamic state (temperature $T$, chemical potentials $\mu_i$ conjugate to conserved charges) become \emph{state functions} ($T \to T(t,\bm{x})$, $\mu_i\to \mu_i(t,\bm{x})$, etc.) that depend on space and time.
Their evolution equations, which involve to the flow velocity, $\UU(t,\bm{x})$, of the plasma or fluid, follow from energy--momentum and charge conservation \citep[Chapter~15]{Landau-vol6}.

Details of the microscopic physics are encoded in \emph{kinetic coefficients} (viscosity,
diffusion coefficient, conductivities, etc.).
Most hydrodynamical systems are ``agnostic'' about the quantum nature of the microscopic degrees of freedom.
Situations where quantum physics affects the hydrodynamical description of the macroscopic systems are rare.
Superfluidity \citep[Chapter~16]{Landau-vol6} is a prominent example of a phenomenon where the quantum nature of the underlying particles drastically changes the properties of the fluid at macroscopic scales.

Two decades ago, a novel phenomenon tied to quantum
physics was identified \citep[see, e.g.,][]{Kharzeev:2011vv,
Kharzeev:12a,Zakharov:2012vv,Giovannini:2013oga,Kharzeev:2013ffa,
Miransky:2015ava,Kharzeev:2015znc}.
The hydrodynamical description of magnetized systems of relativistic fermions in weakly coupled plasmas \citep{Alekseev:98a,Giovannini:2013oga}, of quasi-particles in new materials such as graphene \citep{Miransky:2015ava}, and of the quark--gluon plasma \citep{Kharzeev:2015znc} cannot be formulated in terms of only the standard magnetohydrodynamic (MHD) variables (flow velocity $\UU$, magnetic field $\BB$, density of plasma $\rho$, and pressure $p$) appearing in the Navier-Stokes and the Maxwell equations.
The hydrodynamics of a chiral plasma necessarily contains an additional degree of freedom corresponding to a spacetime-dependent chemical potential, conjugated to the \emph{number difference between right-chiral and left-chiral fermions}.
The dynamics of this degree of freedom is coupled to
the magnetic helicity.

Many new effects arise, among which the most notable one is
the \emph{chiral magnetic effect} (CME), namely the presence of a
contribution to the electric current parallel to the magnetic field.
This effect has first been described by \citet{Vilenkin:80a}
and rederived later using different arguments \citep[see, e.g.,][]
{Redlich:1984md,Tsokos:85,Alekseev:98a,Frohlich:2000en,Frohlich:2002fg,
Fukushima:08,Son:2009tf}.
This contribution to the electric current causes an instability in the system \citep{Joyce:97} that has been analyzed in many works \citep{Frohlich:2000en,Ooguri:2011aa,Boyarsky:11a,Kumar:2014fta,Grabowska:2014efa,Manuel:2015zpa, Buividovich:2015jfa,Boyarsky:15a,Kumar:2016awq,Kumar:2016xuh}.  This instability may be relevant in the physics of the early universe \citep{Joyce:97,Frohlich:2000en,Frohlich:2002fg,Semikoz:04a,Semikoz:2009ye,Boyarsky:11a,
Boyarsky:12a,Semikoz:12a,Tashiro:2012mf,Dvornikov:2011ey,Dvornikov:2012rk,Dvornikov:2013bca,
Dvornikov:2016jth,Manuel:2015zpa,Gorbar:2016klv,Pavlovic:2016mxq,Pavlovic:2016gac},
of the quark-gluon plasmas
\citep{Akamatsu:2013pjd,Taghavi:2013ena,Hirono:2015rla}, or of neutron
stars \citep{Ohnishi:2014uea,Dvornikov:2015lea,Dvornikov:2014uza,
Sigl:2015xva,Yamamoto:2016xtu, Dvornikov:2016cmz}.
However, to the best of our knowledge, a systematic analysis of the system of \emph{chiral MHD} equations, including the back-reaction of the magnetic field on the chiral chemical potential and the coupling to the plasma velocity field, $\UU(t,\bm x)$ appears to be missing.

Our present paper fills this gap.
We derive the system of chiral MHD equations, which involves the magnetic field $\BB$, the fluid velocity field $\UU$, and the chiral chemical potential $\mu_5$.
We analyze magnetic field instabilities (dynamos) described by these equations.
Apart from the laminar dynamo (extensively discussed in the works cited above),
\emph{which will be referred to as the $v_\mu^2$ dynamo}, all other effects
proposed in the present paper:
\begin{asparaitem}[--]
\item the laminar $v_\mu$--shear dynamo;
\item the laminar $v_\mu^2$--shear dynamo;
\item the chiral turbulent $\alpha_\mu$ effect;
\item different kinds of turbulent large-scale dynamos;
\end{asparaitem}
and the modifications of the MHD waves by the CME
appear to be new.
(Here, $v_\mu$ refers to the product of the microscopic Ohmic magnetic diffusion and chiral chemical potential.)
These types of dynamos are so-called \emph{``slow dynamos,''} for which the growth rate tends to zero when the magnetic diffusion due to the electrical conductivity of the plasma tends to zero.

Our paper is organized as follows:
In Section~\ref{sec:CME}, we explain the basic ideas underlying the microscopic origin of the chiral MHD equations, and list some references.
In Section~\ref{sec:governing_eqs}, we consider a complete system of chiral MHD equations, and discuss the astrophysically important limit of small microscopic magnetic diffusion.
In Section~\ref{sec:dynamo-equations} dynamo equations are presented  and conservation laws in chiral MHD are discussed.
In Section~\ref{sec:waves}, we describe MHD waves as modified by the CME.
In Section~\ref{sec:laminar-dynamos}, we study laminar $v_\mu^2$, $v_\mu^2$--shear, and $v_\mu$--shear dynamos, and, in Section~\ref{sec:mean-field-dynamo}, we
discuss the chiral turbulent $\alpha_\mu$ effect and investigate
different kinds of turbulent large-scale dynamos.
In particular, we exhibit turbulent mean-field dynamos with zero mean kinetic helicity involving uniform and nonuniform chiral chemical potentials.
In this section we also analyze possible nonlinearities in turbulent large-scale dynamos.
In particular, we derive an evolution equation for the small-scale
magnetic helicity, which appears as  nonlinearity in mean-field dynamos
with the CME.
In Section~\ref{sec:chiral-mhd-equations}, we derive the chiral MHD equations in an expanding universe.
Finally, in Section~\ref{sec:discussion}, we discuss our results, sketch future studies, and draw some conclusions.

In a separate paper (Part II: numerical simulations~\citep{paper2})
we will investigate the different dynamo effects using numerical simulations, and consider detailed applications of our results to astrophysical systems: the early universe, neutron stars, and the quark--gluon plasma.
Scaling aspects of the inverse turbulent cascade in different regimes
are discussed by \cite{Brandenburg:2017rcb}.

\section{Chiral anomaly and the CME}
\label{sec:CME}

\subsection{Axial symmetry and axial anomaly}
\label{sec:axial-symmetry-anomaly}

In this paper we consider a hydrodynamic description of a plasma consisting of charged massless particles with spin $1/2$, such as a high-temperature electron--positron plasma.
For massless fermions, besides  the conserved electric charge, there is an additional ``classical'' symmetry, generated by the classically conserved \emph{axial charge}, $Q_5$.
This charge counts the number of ``left-chiral'' particles \emph{minus} the number of
``right-chiral'' particles
\citep[see, e.g.][Sections~3.2--3.4]{Peskin-Schroeder}.
In one spatial dimension left/right chiral particles actually travel to the left/to the right, respectively.
There is a classically conserved \emph{axial current} $J^\mu_5 \equiv (c n_5, \JJ_5)$, with the property that
\begin{equation}
  \label{eq:39}
  Q_5 = \int d^3 \bm x \, n_5.
\end{equation}
In a theory of free massless fermions not coupled to any gauge field,
\begin{eqnarray}
Q_5 &=& \text{number of left-chiral particles}
\nonumber\\
&&- \text{number of right-chiral particles} \, .
\end{eqnarray}
The axial current is known to be \emph{anomalous} -- its conservation is destroyed by the presence of gauge fields \citep{Peskin-Schroeder,Treiman:85}.
Therefore the continuity equation of $J^\mu_5$ becomes inhomogeneous
(i.e., it has a source term):
\begin{eqnarray}
  \label{eq:5}
\frac{\partial n_5}{\partial t} + \nab \cdot \JJ_5 = \frac{2e^2}{\pi \hbar^2 c} \EE\cdot \BB ,
\end{eqnarray}
where $e$ is the electric charge, $\hbar$ is Planck's constant, $c$ is the speed of light, $\EE$ is the electric field, $\BB$ is the magnetic field, and $\JJ_5$ is the axial current density.
Here we use Gaussian units, in accordance with the literature in plasma physics and astrophysics.
In some papers, the calculation of the coefficient on the right-hand side (rhs) of
Equation~\eqref{eq:5} is done in the Heaviside-Lorentz system
of units, which would give a coefficient $e^2/(2\pi^2 \hbar^2 c)$.
The presence of $\hbar$ on the rhs of Equation~(\ref{eq:5}) indicates that the inhomogeneity is a quantum effect.
It is possible to add to $J^\mu_5$ a term that would make the new current divergence-free, but it is then no longer gauge-invariant \citep{Frohlich:2000en}.
It is impossible to define an axial current that is simultaneously conserved and
gauge-invariant.
This has consequences for ``real-world observables.''
For example, the neutral meson $\pi^0$ decays extremely fast into two photons, although selection rules based on classical symmetries predict suppression of this decay~\citep{Steinberger:1949wx,Adler:1969gk,Bell:1969ts}.
The phenomenology of the Quantum Hall effect can be derived from the chiral anomaly cancellation \citep[see, e.g., the review by][]{Bieri:2010za}.

By integrating Equation~\eqref{eq:5} over space, we find
\begin{eqnarray}
  \frac{d Q_5}{dt} &=& \frac{2\alphaem}{\pi\hbar}\int d^3 \bm x\,\EE\cdot \BB \nonumber \\
  &=& -\frac{\alphaem}{\pi\,\hbar c} \, \frac{d}{dt} \int d^3\bm{x} \,\AAA \cdot \BB ,
  \label{eq:6}
\end{eqnarray}
where $\AAA$ is the electromagnetic vector potential and $\BB = \nab \times \AAA$.
Here $\alphaem$ denotes the \emph{fine-structure constant}
\begin{eqnarray}
  \label{eq:7}
  \alphaem \equiv \frac{e^2}{\hbar c} \approx {1 \over 137} .
\end{eqnarray}
Thus, quantum-mechanically, the variation of the axial charge in time is proportional to the  variation of the \emph{magnetic helicity} defined by
\begin{eqnarray}
  \label{eq:8}
  \helicity \equiv \int d^3\bm{x} \,\AAA \cdot \BB .
\end{eqnarray}
The magnetic helicity is gauge invariant, provided that $\BB$ is parallel to the boundary of the integration domain.
Under the assumption that the magnetic field vanishes at infinity, the definition~\eqref{eq:8} of $\helicity$ coincides with the standard definition of magnetic helicity.
Using Equation~\eqref{eq:6}, we can define a new charge, $Q_5 +(\alphaem/\pi\hbar c) \, \helicity$~\citep[see][]{Alekseev:98a}, which for \emph{massless} fermions is conserved, i.e.,
\begin{eqnarray}
  \label{eq:9}
\frac{d}{dt} \left( Q_5 +\frac{\alphaem}{\pi\hbar c}\helicity\right) = 0 .
\end{eqnarray}
The axial anomaly implies that,
although this  quantity is conserved, it is not possible to construct a corresponding \emph{gauge-invariant} current.
Equation~\eqref{eq:9} shows that, by changing the magnetic
helicity, one can create or destroy chirality in the fermion state of the system.
Vice versa, a change in the occupation of right- and left-chiral fermions
leads to generation or decay of magnetic helicity.
This has drastic consequences in MHD, as shown below.

As a consequence of the axial anomaly (and of the nonconservation of
the axial charge $Q_5$), there is an additional term in the \emph{electric
current} proportional to the \emph{chiral chemical potential} $\mu_5$,
conjugated to the axial charge, in the usual thermodynamical sense.
This electric current due to the chiral magnetic effect (CME) is given by
\begin{eqnarray}
  \label{eq:13}
   \JJ_{\rm CME}
  = \frac{\alphaem}{\pi \hbar} \mu_5 \BB ,
\end{eqnarray}
where $\mu_5$ is the chemical potential conjugated to the conserved charge~\eqref{eq:9} \citep{Alekseev:98a,Frohlich:2000en}. For completeness we sketch a simple derivation of Equation~\eqref{eq:13} in Appendix~\ref{sec:CME_current_LLL}.

\subsection{Inhomogeneous chiral chemical potential}
\label{sec:current-with-theta}

Equation~\eqref{eq:13}
remains valid when the magnetic field $\BB(t,\bm{x})$ and the chiral chemical potential $\mu_5(t,\bm{x})$ depend on space and time.  However, for a spatially inhomogeneous chemical potential, additional terms appear, expressing the relaxation of the inhomogeneous chemical potential.
This question has been first explored by \citet{Frohlich:2000en,Frohlich:2002fg}.
Later, it also appeared in the context of a quark--gluon plasma \citep{Kharzeev:2007tn,Kharzeev:2009fn,Ozonder:2010zy,Kharzeev:2011rw,Kalaydzhyan:2012ut, Zhitnitsky:2012im,Zhitnitsky:2013hs,Landsteiner:2012kd,Huang:2013iia}.
The treatment of the inhomogeneous chiral chemical potential in the hydrodynamical approach can be found in~\cite{Kharzeev:11a}. We briefly summarize the main ideas below.

The inhomogeneous chiral chemical potential leads to the appearance of a local electric charge density.
Instead of expressing the current as a complicated nonlocal function of
$\mu_5(t,\bm{x})$ and of the electric charge density $\varrho(t,\bm{x})$,
it is convenient to introduce a field $\Theta(t,\bm{x})$ with the property that the electric charge density induced by $\Theta$ is given by
\begin{eqnarray}
  \label{eq:29}
  \varrho_\Theta \equiv - \BB\cdot \nab\Theta ,
\end{eqnarray}
and the chiral chemical potential is
\begin{eqnarray}
  \label{eq:85}
  \mu_5 \equiv \pfrac{\Theta}{t} \frac{\hbar\pi}{\alphaem} .
\end{eqnarray}
The generalization of the current given by Equation~\eqref{eq:13} to the inhomogeneous case can be expressed in terms of $\Theta$:
\begin{eqnarray}
  \label{eq:3}
   \JJ_{\rm CME}
\to \JJ_\Theta = \BB\pfrac{\Theta}{t} + c \nab\Theta \btimes \EE .
\end{eqnarray}
The first term on the rhs of Equation~\eqref{eq:3} is known from Equation~\eqref{eq:13}.
The second term shows that if the gradient of $\Theta$ does not vanish, an additional term  appears in the current. It is perpendicular to the \emph{electric} field.
This term induces a spatial variation of the electric charge density
(see Equation~\eqref{eq:29}), and hence, additional currents may appear.

There are two important properties of the current given by Equation~\eqref{eq:3}:
\begin{enumerate}
\item It is \emph{dissipationless}, i.e.,\ it does not generate entropy,
and the corresponding kinetic coefficient is \emph{even} with respect
to time reversal.
\item It is conserved by itself (without invoking further contributions to the electric current and/or equations of motion).
This property can be checked explicitly by calculating the divergence $\nab \cdot \JJ_\Theta$ and noticing that
\begin{eqnarray}
  \label{eq:4}
  \nab \cdot \JJ_\Theta  = \pfrac{}{t}\Bigl(\nab\Theta \cdot \BB\Bigr) ,
\end{eqnarray}
so that, in view of Equation~\eqref{eq:29}, we get
\begin{eqnarray}
  \label{eq:30}
  \pfrac{\varrho_\Theta}{t} + \nab\cdot \JJ_\Theta = 0 .
\end{eqnarray}
\end{enumerate}
This can also be seen by introducing the 4-current
\begin{eqnarray}
  \label{eq:86}
  J_\Theta^\mu \equiv (c\varrho_\Theta,\JJ_\Theta) ,
\end{eqnarray}
and noticing that Equations~\eqref{eq:29} and \eqref{eq:3} can be combined to
\begin{eqnarray}
  \label{eq:32}
J_\Theta^\mu = \frac c2 \epsilon^{\mu\nu\lambda\rho}\partial_\nu\Theta F_{\lambda\rho} .
\end{eqnarray}
Thus,
\begin{equation}
  \pfrac{J_\Theta^\mu}{x^\mu} = \frac c2 \epsilon^{\mu\nu\lambda\rho}(\partial_\mu \partial_\nu\Theta)F_{\lambda\rho} + \frac c2 (\partial_\nu\Theta) \epsilon^{\mu\nu\lambda\rho} \partial_\mu F_{\lambda\rho} = 0 ,
  \label{eq:36}
\end{equation}
where the first term on the rhs is zero because it is a contraction of the
symmetric $(\partial_\mu \partial_\nu\Theta)$ tensor with the antisymmetric one, while the second term is just a Bianchi identity.

\section{Equations of chiral MHD}
\label{sec:governing_eqs}

In this paper we consider a one-fluid MHD model
that follows from two-fluid model of plasma
(for details we refer the reader to different books on plasma physics:
\citep{Artsimovich-Sagdeev,Melrose-QuantumPlasmadynamicsMagnetized,Biskamp:97}).
In this section we present the complete set of \emph{chiral MHD equations}, including the field equation for an inhomogeneous chiral chemical potential.

\subsection{The Maxwell equations}
\label{sec:mhd-equations}

The system of MHD equations consists of the Maxwell equations,

\begin{subequations}
  \label{eq:maxwell}
  \begin{eqnarray}
    \nab \times \BB &= &{4 \pi \over c} \JJ_\tot + {1\over c}{\partial\EE \over \partial t} ,
    \label{rotB}\\
    \nab \cdot \EE&=& 4\pi \varrho_\tot ,
    \label{eq:80}\\
    \nab  \times   {\EE}&=& - \frac{1}{c} \frac{\partial \BB}{\partial t},                         \label{rotE} \\
    \nab\cdot \BB &=& 0 ,
    \label{divB1}
  \end{eqnarray}
\end{subequations}
and the Navier-Stokes equation, describing the evolution of the velocity field, $\UU$
(see Sections~\ref{sec:one-component-relativistic} and~\ref{sec:two-component}).
The velocity field $\UU$ is a weighted average of velocities of individual
microscopic components \citep[Sections~2.4--2.5]{Artsimovich-Sagdeev}.

\subsection{Electric currents}
\label{sec:electric-current}

The matter current $J^\mu_\tot \equiv (c\varrho_\tot,\JJ_\tot)$ consists of
several different terms.
The most obvious one is related to the charge density in the plasma,
\begin{equation}
  \label{eq:62}
  J^\mu = \varrho_\el u^\mu = \gamma(\rho_\el,\rho_\el\UU) ,
\end{equation}
where $u_\mu$ is the 4-velocity vector field, $u^\mu = \gamma(c, \UU)$,
and $\gamma = (1-\UU^2/c^2)^{-1/2}$ is the Lorentz factor.
There is also an Ohmic current \citep{Melrose-QuantumPlasmadynamicsMagnetized}
\begin{equation}
  \label{eq:59}
  J^\mu_\ohm = \sigma \gamma F^{\mu\nu}u_\nu ,
\end{equation}
where $\sigma$ is the electrical conductivity of the fluid.
In this paper we only consider temporal and spatial scales with the property that $\sigma(\omega,\bm k) \approx \sigma$ and hence $|\UU|\ll c$, $\gamma \approx 1$.
In this regime the spatial component of the Ohmic current
given by Equation~\eqref{eq:59} is
\begin{equation}
\JJ_\ohm = \sigma \left({\EE} + \frac{1}{c} \UU  \times   {\BB} \right) .
 \label{JJ}
\end{equation}
The total charge density is defined as
\begin{equation}
  \label{eq:76}
  \varrho_\tot \equiv J^\mu_\tot \cdot u_\mu \,,
\end{equation}
and electric neutrality of the plasma on distance scales described by
hydrodynamic equations implies that
\begin{equation}
  \label{eq:21}
  \varrho_\tot = 0 .
\end{equation}
Given that $u_\mu J^\mu_\ohm = 0$, for any configuration of fields
and velocities, the electric neutrality condition of
\emph{conventional MHD} in the absence of a CME
implies that $\varrho_\el = 0$, i.e., the
current given by Equation~\eqref{eq:62} vanishes, so that the total current is
given by the Ohmic current.

The situation is more interesting in \emph{chiral MHD}.
Here the total electric current, $\JJ_\tot$, and the total charge
density, $\varrho_\tot$, acquire additional contributions proportional to
the extra field $\Theta(t,\bm x)$, as expressed  by Equation~\eqref{eq:32}.
The total current is the sum of the longitudinal contribution (Equation~\eqref{eq:62}),
the Ohmic (transversal) contribution (Equation~\eqref{eq:59}),
and the chiral contribution (Equation~\eqref{eq:32}):
\begin{equation}
  \label{eq:79}
  J^\mu_\tot = \varrho_\el u^\mu + J^\mu_\ohm + \frac c2 \epsilon^{\mu\nu\lambda\rho}\partial_\nu\Theta F_{\lambda\rho} .
\end{equation}
The total electric charge density~\eqref{eq:76} receives an additional contribution proportional to $\nab\Theta$:
 \begin{equation}
  \label{eq:2}
  \varrho_\tot  =  \varrho_\el-\BB\cdot \nab \Theta\;.
\end{equation}
The neutrality condition~\eqref{eq:21}
allows us to express $\varrho_\el$ by
\begin{equation}
  \label{eq:89}
  \varrho_\el = \BB\cdot\nab\Theta .
\end{equation}
The total electric current $\JJ_\tot$ (Equation~\eqref{eq:79}) is given by
\begin{eqnarray}
  \JJ_\tot \equiv \varrho_\el \UU + \JJ_\ohm + \JJ_\Theta .
  \label{eq:34}
\end{eqnarray}
Using Equation~\eqref{eq:89}, we obtain the explicit expression for
the total current:
\begin{eqnarray}
  \label{eq:92}
  \JJ_\tot &=&  (\BB\cdot\nab\Theta)\UU + \sigma\biggl(\EE + \frac1c{\UU\times \BB}\biggr) \nonumber \\
  & & +   \biggl(\dot\Theta \BB + c \, \nab\Theta \times \EE\biggr) .
\end{eqnarray}

Using the identity $\UU (\BB\cdot\nab\Theta) = \BB(\UU\cdot\nab\Theta) +\nab\Theta\times(\UU\times \BB)$, we rewrite the total electric current in the form
\begin{eqnarray}
  \label{eq:93}
  \JJ_\tot &=& \sigma\biggl(\EE + \frac1c\UU\times \BB\biggr) \nonumber \\
           & & + \frac{D\Theta}{Dt} \BB  + c \, \nab\Theta \times \left(\EE + \frac{1}c \UU\times \BB\right) ,
\end{eqnarray}
where the \emph{advective time derivative} of $\Theta$
is defined by
\begin{equation}
  \label{eq:23}
  \frac{D\Theta}{Dt} \equiv \pfrac\Theta t + \UU\cdot \nab\Theta .
\end{equation}

\subsection{Electric field for small magnetic diffusion}
\label{sec:closing-system}

Conventionally, MHD is formulated as the evolution of the magnetic and the velocity field,
neglecting the Maxwell displacement current in the Maxwell equation~\eqref{rotB}.
Substituting the expression~\eqref{eq:93} into Equation~\eqref{rotB},
we can express the electric field $\EE$
in terms of $\BB$, $\Theta$, and $\UU$ by
\begin{eqnarray}
  \label{eq:94a}
  \EE &=& - \frac{1}{c} \, \biggl[\UU\times \BB - \eta \left(\nab\times \BB -
  \mu \BB\right)
  \nonumber \\
  & & + 4 \pi\eta \, \nab\Theta \times\left(\EE + \frac{\UU\times \BB}c \right) \biggr] ,
\end{eqnarray}
where
\begin{equation}
  \label{eq:47}
  \eta\equiv \frac{c^2}{4 \pi \sigma}
\end{equation}
is the magnetic diffusion coefficient due to electrical conductivity of the fluid, and $\mu$ is defined by
\begin{equation}
\mu \equiv {4 \pi \over c} \frac{D \Theta}{D t}
 = {4 \alphaem\over \hbar c} \mu_5 .
   \label{theta}
\end{equation}

Let us consider the special case of small magnetic diffusivity ($\eta \to 0$) typical for astrophysical systems with large magnetic Reynolds numbers.
More precisely, we consider a fluid with large Reynolds number,
$Re=V L/\nu \gg 1$ (the ratio of the nonlinear term
to the viscous term in the momentum equation), and large magnetic
Reynolds number $Rm=V L/\eta \gg 1$
(the ratio of the nonlinear term to the magnetic diffusion term in the induction equation),
where $V$ and $L$ are characteristic velocity and length scales
of the system, respectively, and $\nu$ is the kinematic viscosity.
For large magnetic Reynolds numbers, we can neglect the terms
of order $\sim O(\eta^2)$, which yields
\begin{eqnarray}
  \label{eq:94}
  \EE &=& - \frac{1}{c} \, \biggl[\UU\times \BB - \eta \left(\nab\times \BB
 - \mu \BB\right)
  \nonumber\\
  &&+ \frac{4 \pi\eta^2}{c} \, \nab\Theta \times\left(\nab\times \BB
 - \mu \BB\right) \biggr].
\end{eqnarray}
In this equation,
the $\nab\Theta$ terms only appear in second order in $\eta$.
If one only keeps first-order terms in $\eta$, the $\nab\Theta$ terms
can be dropped.
This is valid when $(4 \pi \eta/c) \, |\nab\Theta| \ll 1$.

\subsection{Dynamic equation for the chiral chemical potential}
\label{sec:chiral-mhd}

The extra term, $\JJ_\Theta$, in the electric current depends on a new field, $\Theta(t,\bm x)$.
The details of the evolution of $\Theta(t,\bm x)$ depend on the underlying microscopic model.
In this paper, we assume that the dissipation of $\mu_5$ is determined by an inhomogeneous diffusion equation of the form
\begin{equation}
\frac{D \mu_5}{D t} = D_5 \, \Delta \mu_5 +\Lambda^{-2}
{\EE} {\bm \cdot} {\BB},
\label{mu}
\end{equation}
where $D_5$ is a diffusion coefficient.
Equation~\eqref{mu} expresses the dynamical law of the new field
$\Theta$.
Combining Equation~\eqref{mu} with Equation~\eqref{theta}, we see that the parameter
$\Lambda$ in Equation~\eqref{mu} is given by \citep{Boyarsky:15a}
\begin{eqnarray}
\Lambda^{-2} \equiv \frac{12 \alphaem}\pi\, \frac{\hbar^2 c^3}{k_B^2 T^2} .
\label{eq:81}
\end{eqnarray}
It is an (inverse) \emph{susceptibility} (i.e., the response
of the axial charge to a change in the chiral chemical potential).
In expression~\eqref{eq:81},
$T$ is the temperature and $k_{\rm B}$ is the Boltzmann constant.
Note that different choices of evolution equations for
$\mu_5$ are possible, depending on the microscopic physics;
see the discussion in \cite{Boyarsky:15a}.
In particular, an evolution equation for $\mu_5$ with a damping term,
$- \Gamma_f \mu_5$, and a source term, $\Gamma_{\rm sr} \mu_5$, has recently been used
by \citet{Pavlovic:2016gac} to study the influence of the chiral
anomaly on the evolution of MHD turbulence. Here
$\Gamma_f$ is the total chirality flipping rate, and $\Gamma_{\rm
sr}$ is a source term that takes into account the generation of
chiral asymmetry.

\subsection{Underlying models of matter}
\label{sec:underlying-matter}

Before discussing the equations of motion for the fluid by specifying
the dynamical equation for $\UU$, we discuss some microscopic models of matter
that appear to lead to the chiral MHD equations.

Our choice of electric currents considered in
Section~\ref{sec:electric-current} refers to a certain physical models.
They consist of a nonrelativistic plasma whose electric properties
are described by the current $\JJ_\ohm$ and electric charge density
$\varrho_\el$, which may not vanish.

The nonrelativistic dynamics of the plasma is governed by the Maxwell equations
and the Navier-Stokes  equation (relating the fluid velocity, $\UU \ll c$
to the magnetic field, $\BB$).
The nonrelativistic plasma interacts with a highly relativistic plasma component.
The electric current, $\JJ_\Theta$, caused by the
relativistic plasma component, is an additional source for the magnetic field
in the Maxwell equations [see Equations~\eqref{rotB} and \eqref{eq:34}].
Such plasmas arise in the description of certain astrophysical systems,
where a nonrelativistic plasma interacts with cosmic rays consisting of
relativistic particles with small number density;
see, e.g., \cite{Schlickeiser-book}.
The cosmic-ray current may trigger the ``Bell
instability'' \citep{Bell:04,Bykov:11}, which produces helical turbulence,
and the coupling of helical turbulence with the cosmic-ray
current results in the generation of large-scale magnetic fields by
a mean-field dynamo action \citep{Rogachevskii:2012ah}.

One can envisage a different model of matter in which all charged
particles are highly relativistic.
The relativistic velocities of the particles do not imply, however,
that the hydrodynamic bulk velocity $\UU$ is relativistic.
An example of a physical situation in which this system is realized, is a
plasma of hot relativistic particles, as it arises in the early Universe.
The corresponding equations of relativistic MHD can be derived from
the covariant energy--momentum conservation on the background of a
Friedmann--Lema\^itre universe.
The 4-velocity of the fluid contains a contribution from isotropic Hubble expansion and additional terms caused by magnetic fields present in the plasma, which
can be treated consistently
(the fluid velocity is much smaller than the speed of light).
The resulting equations describing the evolution of the magnetic field and
of these particular velocities can be reduced to the coupled
a Navier--Stokes--Maxwell equations (in comoving coordinates).
This approach is discussed in detail in
\citep{Jedamzik:96,Brandenburg:96,Banerjee:04} and in reviews of the
subject \citep{Barrow:2006ch,BS05,Subramanian:09,Durrer:13}.
In Section~\ref{sec:chiral-mhd-equations} we demonstrate that this framework
remains intact if one adds the current $\JJ_\Theta$ and the evolution of the
$\Theta$ field in the equations.

\subsection{Equation of motion for one-component relativistic plasmas}
\label{sec:one-component-relativistic}

In this section we consider an equation of motion for the
one-component relativistic plasma:
\begin{eqnarray}
\rho{D \UU \over D t}&=& \frac{1}{c} \,\JJ_\tot  \times   \BB
-\nab p  + \nab  {\bm \cdot} (2\nu \rho \SSSS)
+\rho \ff ,
\label{UU-DNS0}
\end{eqnarray}
where $\rho$ is the mass density of the plasma, $p$ is the plasma pressure,
$\rho \ff$ is an external force, and
\begin{equation}
  \label{eq:53}
  {\sf S}_{ij}\equiv\frac12(U_{i,j}+U_{j,i})-\frac13 \delta_{ij} \nab
  {\bm \cdot} \UU
\end{equation}
is the traceless strain tensor (commas denote partial spatial derivatives).
The magnetic field affects the dynamics of the velocity field in the
Navier-Stokes equation~\eqref{UU-DNS0} via the \emph{Lorentz force},
$c^{-1} \, \JJ_\tot \times \BB$, which
necessarily  contains the \emph{total} electric current, $\JJ_\tot$,
regardless of its origin (see Appendix~\ref{sec:lorentz-force} for details).

We now take into account that, using Equation~(\ref{rotB}),
\begin{eqnarray}
&& \frac{1}{c} \,\JJ_\tot  \times   \BB
= \frac{1}{4\pi} \,(\nab   \times   {\BB})  \times  \BB ,
\label{Lor-force-1}
\end{eqnarray}
where we have neglected the displacement current
$c^{-1} \partial\EE/\partial t$.
Thus, the Navier-Stokes equation is given by
\begin{eqnarray}
\rho{D \UU \over D t}&=& \frac{1}{4\pi} \,(\nab   \times   {\BB})  \times   \BB
-\nab p + \nab  {\bm \cdot} (2\nu \rho \SSSS)
+\rho \ff .
\nonumber\\
\label{UU-DNS1}
\end{eqnarray}

In this paper we focus our attention on an \emph{isothermal fluid:} $T=\mathrm{const}$.
In principle, the temperature $T$ should also be treated as a field $T(\bm x,t)$
determined by an entropy equation
(including Ohmic dissipation, radiation, etc.) that determines its evolution.
The presence of a $\Theta$-field may introduce additional terms in the temperature equation.
However, the equilibration rate of the temperature gradients is related to the shortest timescales of the plasma (of the order of the plasma frequency or below) and is much shorter than the
time-scales that we consider in the present study.
Therefore, the isothermal approximation is consistent, and we leave a more general treatment for future work.
In Section~\ref{sec:chiral-mhd-equations} we comment on the applicability of this assumption
in studies of the early universe, where temperature is a function of time.

\subsection{Equation of motion for two-component relativistic and nonrelativistic plasmas}
\label{sec:two-component}

In a nonrelativistic plasma of relativistic particles,
the Navier-Stokes equation for the plasma motion reads
\begin{equation}
  \label{eq:75}
\rho  \frac{D\UU}{Dt} = -\nab p + \frac{1}{c} \, \JJ_\ohm \times \BB+ \nab  {\bm \cdot} (2\nu \rho \SSSS) +\rho \ff,
\end{equation}
where $\UU$ is the velocity field of the non-relativistic plasma
and $\JJ_\ohm$ is the Ohmic current.
We now take into account that
\begin{eqnarray}
&& \frac{1}{c} \, \JJ_\ohm  \times   \BB =  \frac{1}{4 \pi} \ (\nab   \times   {\BB})  \times  \BB - {\EE} ({\BB} {\bm \cdot} \nab )\Theta
\nonumber \\
\quad  & & + ({\EE} {\bm \cdot} {\BB}) (\nab  \Theta)
 =  \frac{1}{4 \pi} \, (\nab   \times   {\BB})  \times  \BB + \mathcal{O}(\eta^2),
\label{Lor-force}
\end{eqnarray}
where we used Equations~(\ref{eq:3}), (\ref{JJ}), and~(\ref{eq:34}) for  $\varrho_\tot=0$.
As was shown in Section~\ref{sec:closing-system},
the terms proportional to $\nab\Theta$  only appear in second order in $\eta$
in the expression for the electric field [see Equation~(\ref{eq:94})].
Since, in the present study, we only consider plasmas with large
magnetic Reynolds numbers, we only
keep first-order terms in $\eta$, and the $\nab\Theta$ terms
in the expression for the electric field can be dropped.
Substituting Equation~(\ref{Lor-force}) into Equation~(\ref{eq:75}), we
obtain the Navier-Stokes equation for a nonrelativistic plasma:
\begin{eqnarray}
\rho{D \UU \over D t} &=& \frac{1}{4 \pi} \,(\nab   \times   {\BB})  \times   \BB
-\nab  p + \nab  {\bm \cdot} (2\nu \rho \SSSS) \nonumber \\
  & &+\rho \ff + O(\eta^2) .
\label{UU-DNSS}
\end{eqnarray}

Note that, for two-component relativistic and nonrelativistic plasmas,
the total current is $\JJ_\tot=\JJ_\ohm +\JJ_\Theta + \JJ_{\rm rel}$,
where $\JJ_{\rm rel}$ is the Ohmic current caused by the relativistic plasma component
and $\JJ_\Theta$ is the current that caused by the CME.
To highlight the effect of $\JJ_\Theta$, we assume here
that the Ohmic component, corresponding to the current $\JJ_{\rm rel}$, caused by
the relativistic plasma is much smaller than
the current $\JJ_\Theta$.
If the current $\JJ_{\rm rel}$ is not small,
an additional term, $-c^{-1}\, \JJ_{\rm rel}  \times   \BB$, on
the rhs of Equation~(\ref{UU-DNSS}) appears.
This term can cause a small-scale Bell instability~\citep{Bell:04},
production of a helical small-scale turbulence,
and the generation of a large-scale magnetic field~\citep{Rogachevskii:2012ah}.
In this paper these effects are not studied.

To arrive at a consistent description of two-component relativistic and nonrelativistic plasmas, one should supplement
Equation~\eqref{UU-DNSS} with the momentum equation that describes the evolution
of the relativistic plasma component.
In particular, this equation describes interactions of the relativistic
and nonrelativistic plasma components resulting in energy dissipation of the relativistic particles.
The timescale of the dissipation \citep{Schlickeiser-book,Bykov-Brand-review13} is, however, much longer than characteristic timescales of the dynamo instabilities
analyzed below (see also the comment in the last paragraph of Sect.~\ref{sec:one-component-relativistic} about the isothermal assumption).

\section{Dynamo equations in chiral MHD}
\label{sec:dynamo-equations}

In this section we derive the generalized induction equation taking into
account the CME and formulate the dynamo equations in chiral MHD.

\subsection{Induction equations in chiral MHD}

Substituting Equation~(\ref{eq:94}) into Equation~(\ref{rotE}) we obtain the generalized
induction equation:
\begin{eqnarray}
\frac{\partial \BB}{\partial t} &=&
 \nab   \times  \left[{\UU}  \times   {\BB}
- \eta \, \left(\nab   \times   {\BB}
- \mu {\BB} \right)\right]
+ O(\eta^2) .
\nonumber\\
\label{induction-eq}
\end{eqnarray}
The term $\propto \mu {\BB}$ in the induction equation~(\ref{induction-eq}) describes the CME.
Furthermore, expression~(\ref{eq:94}) for the electric field yields
an expression for ${\EE} {\bm \cdot} {\BB}$, which is the source for
evolution of the chiral chemical potential:
\begin{eqnarray}
{\EE} {\bm \cdot} {\BB} &=& \frac{\eta}{c} \left[{\BB} {\bm \cdot} (\nab \times {\BB}) -
\mu {\BB}^2 \right]
+ O(\eta^2) .
\label{EB}
\end{eqnarray}

\subsection{System of dynamo equations}

Using Equations~(\ref{induction-eq}) and (\ref{EB}), we find
the system of chiral MHD equations that
includes the induction equation for the magnetic field $\BB$,
the Navier-Stokes equation for the velocity field $\UU$,
the continuity equation for the plasma density $\rho$,
and the evolutionary equation for the normalized chiral chemical
potential,
$\mu = (4 \alphaem /\hbar c) \mu_5$:
\begin{eqnarray}
\frac{\partial \BB}{\partial t} &=&
\nab   \times   \left[{\UU}  \times   {\BB}
- \eta \, \left(\nab   \times   {\BB}
- \mu {\BB} \right) \right] ,
\label{ind-DNS}\\
\rho{D \UU \over D t}&=& (\nab   \times   {\BB})  \times   \BB
-\nab  p + \nab  {\bm \cdot} (2\nu \rho \SSSS)
+\rho \ff ,
\label{UU-DNS}\\
\frac{D \rho}{D t} &=& - \rho \, \nab  \cdot \UU ,
\label{rho-DNS}\\
\frac{D \mu}{D t} &=& D_5 \, \Delta \mu
+ \lambda \, \eta \, \left[{\BB} {\bm \cdot} (\nab   \times   {\BB})
- \mu {\BB}^2\right],
\label{mu-DNS}
\end{eqnarray}
where
the magnetic field $\BB$ is normalized such that the magnetic energy
density is $\BB^2/2$ without the $4\pi$ factor.\footnote{This is
equivalent to using the Heaviside-Lorentz system of units, except
that in Equation~\eqref{eq:47} the $4\pi$ factor is retained.}
The chiral feedback parameter is
\begin{eqnarray}
\lambda=3 \hbar c \left({8 \alphaem \over k_B T} \right)^2,
\end{eqnarray}
$p=c_{\rm s}^2 \rho$ is the fluid pressure, and $c_{\rm s}$ is the
isothermal sound speed.
In Equations~(\ref{ind-DNS})--(\ref{mu-DNS}),
$\mu$ has the dimension of inverse length, and
$\lambda^{-1}$ has the dimension of energy per
unit length.

\subsection{Conservation law}

Equations~(\ref{ind-DNS}) and~(\ref{mu-DNS}) give rise to a conservation law
that we discuss below.
We use the induction equation and the equation for the vector potential:
\begin{eqnarray}
&&\frac{\partial \BB}{\partial t} = -
\nab   \times   {\bm \EE} ,
\label{ind-2}\\
&&\frac{\partial {\bm A}}{\partial t} = - {\bm \EE} + \nab  \Phi ,
\label{AA-2}
\end{eqnarray}
where $\BB = \nab   \times   {\bm A}$ and $\Phi$ is the electrostatic potential.
Multiplying Equation (\ref{ind-2}) by ${\bm A}$ and Equation (\ref{AA-2}) by $\BB$, and adding them, we
obtain an evolution equation for the magnetic helicity density,
${\bm A} {\bm \cdot} \BB$:
\begin{eqnarray}
&&\frac{\partial {\bm A} {\bm \cdot} \BB}{\partial t} + \nab  {\bm \cdot}
\left({\bm \EE} \times   {\bm A} + \BB \, \Phi\right) = - 2 {\bm \EE} {\bm \cdot} \BB .
\label{AB}
\end{eqnarray}
Since ${\bm \EE} {\bm \cdot} \BB \propto \eta$ (see Equation~(\ref{EB})),
the density of magnetic helicity, ${\bm A} {\bm \cdot} \BB$ , is conserved
in the limit $\eta \to 0$.
Equation~(\ref{mu-DNS}) can be rewritten in the form
\begin{eqnarray}
\frac{\partial (2 \mu/\lambda)}{\partial t} + \nab  {\bm \cdot}
\big[- (2D_5/\lambda) \,\nab  \mu\big] = 2 {\EE} {\bm \cdot} {\BB} ,
\label{mu-2}
\end{eqnarray}
where we have assumed that $D_5$ and $\lambda$ are constant.

Adding Equations~(\ref{AB}) and~(\ref{mu-2}), we find the conservation law:
\begin{equation}
\frac{\partial }{\partial t} \left({\lambda \over 2}  {\bm A} {\bm \cdot} \BB
+ \mu \right) + \nab  {\bm \cdot}
\left[{\lambda \over 2} \left({\bm \EE} \times   {\bm A} + \BB \, \Phi\right)
-  D_5 \nab  \mu \right] = 0 .
\label{CL}
\end{equation}
Thus,
$\half \lambda {\bm A} {\bm \cdot} \BB + \mu$
is conserved for arbitrary $\eta$.
This implies that when ${\bm A} {\bm \cdot} \BB$ increases, during the dynamo action,
the chiral chemical potential $\mu$ must decrease.
We will see that this property is responsible for the dynamo saturation
in the nonlinear stage of evolution of the magnetic field.

\section{Waves in chiral MHD}
\label{sec:waves}

In this section we study the modification of the MHD waves by the CME.
Let us consider the following equilibrium: $\BB_{\rm eq}\equiv\BB_0=\mathrm{const}$, ${\UU}_{\rm eq}=0$,
$\rho_{\rm eq}\equiv\rho_0=\mathrm{const}$ and
$\mu_{\rm eq}\equiv \mu_0 =\mathrm{const}$.

The linearized equations~(\ref{ind-DNS})--(\ref{rho-DNS}) for perturbations
of the plasma density, velocity, and magnetic fields yield
\begin{eqnarray}
 \frac{\partial \BB}{\partial t} &=& \left({\BB}_0 {\bm \cdot} \nab \right) {\UU}
+v_\mu \, \nab   \times   \BB - {\BB}_0 \left(\nab  {\bm \cdot} \UU\right)
+ \eta \Delta \BB,
\nonumber\\
\label{W3}\\
 \rho_0 \frac{\partial \UU}{\partial t} &=& \left({\BB}_0 {\bm \cdot} \nab \right) {\BB} - \nab  \left(p + {\BB}_0 {\bm \cdot} \BB\right) ,
\label{W4}\\
 \frac{\partial \rho}{\partial t} &=& - \rho_0 \left(\nab  {\bm \cdot} \UU\right),
\label{W5}
\end{eqnarray}
where $v_\mu=\eta \, \mu_{0}$ is the velocity caused by the CME.
In Equation~(\ref{W4}) we have omitted for simplicity the damping effects caused
by the kinematic viscosity.
We seek a solution of Equations~(\ref{W3})--(\ref{W5}) in the form
$\BB, \UU, \rho \propto \exp(\tilde \gamma t + i {\bm k} {\bm \cdot} {\bm r})$,
where $\tilde \gamma = \gamma - i \omega$.

\subsection{Incompressible flow}

For an incompressible fluid, Equations~(\ref{W3})--(\ref{W5}) yield
\begin{eqnarray}
 \tilde \gamma_{1,2} &=& -{1 \over 2} (\geta-\gmu) \pm \left[{1 \over 4} (\geta-\gmu)^2 - \oA^2\right]^{1/2},
\label{WN1}\\
 \tilde \gamma_{3,4} &=&  -{1 \over 2} (\geta+\gmu) \pm \left[{1 \over 4} (\geta+\gmu)^2 - \oA^2\right]^{1/2},
\label{WN2}
\end{eqnarray}
where $\oA ={\bm k} \cdot {\bm v}_A$ is the frequency of the Alfv\'{e}n
waves in the absence of the CME, ${\bm v}_A= \BB_0/\sqrt{\rho}$ is the
Alfv\'{e}n speed, $\geta=\eta k^2$, and $\gmu=v_\mu k$.
When $\oA^2 > (\geta  \pm \gmu)^2/4$, there are
Alfv\'{e}n waves with frequency
\begin{eqnarray}
\omega = \pm \left[\oA^2 - {1 \over 4} (v_\mu k \pm \eta k^2)^2 \right]^{1/2} .
\label{W1}
\end{eqnarray}
Equation~(\ref{W1}) implies that the frequency of the Alfv\'{e}n wave
decreases if the CME for an incompressible fluid is taken into account.
The CME can also cause various instabilities or damping of waves
(see next sections).

\subsection{Compressible fluids}

\begin{figure}[!t]
  \centering
  \includegraphics[width=0.5\textwidth]{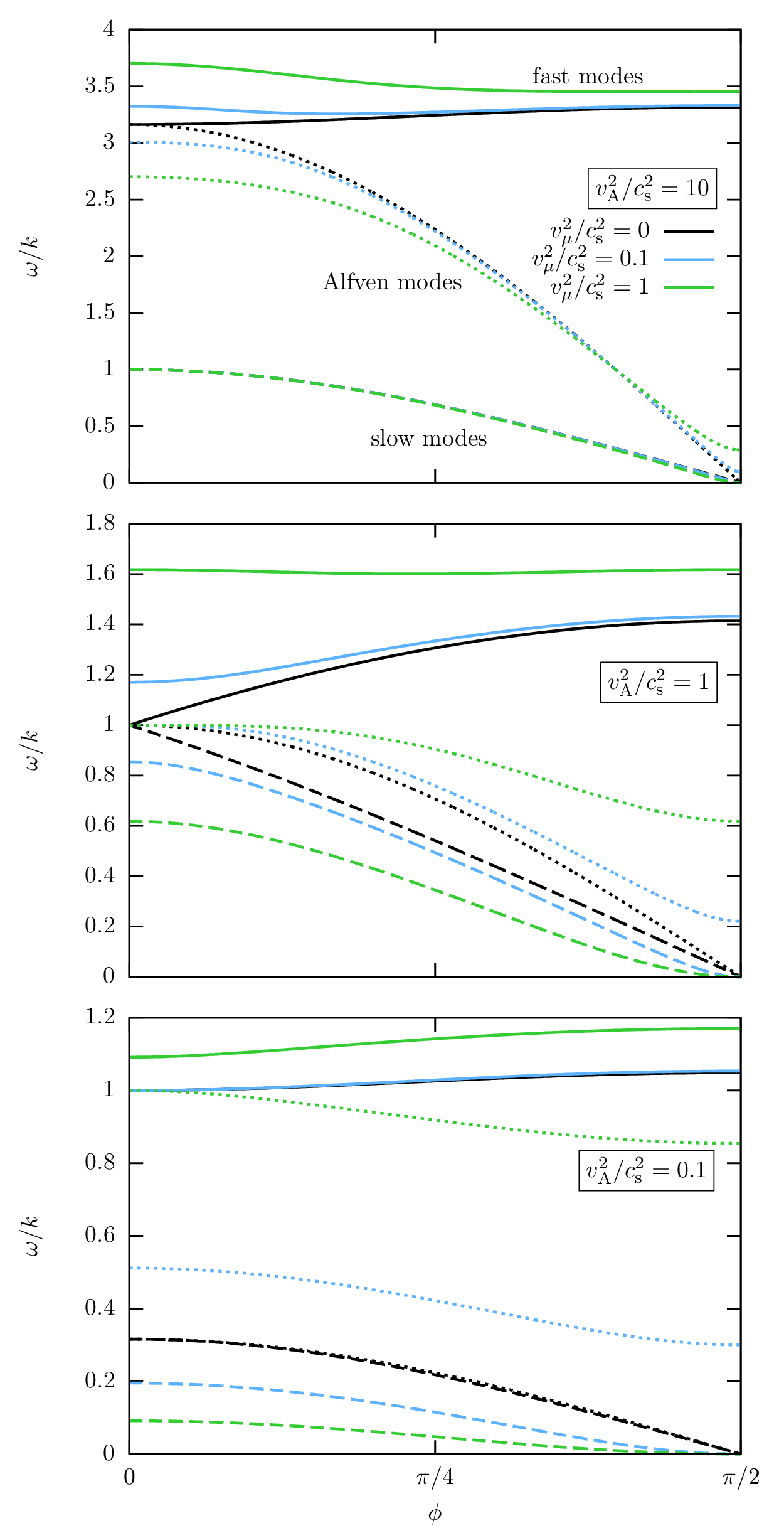}
  \caption{
Influence of the chiral magnetic effect on the MHD waves in a compressible flow.
Shown are the ratios $\omega/k$ vs.\ the angle $\phi$ between the wavevector ${\bm k}$
and the equilibrium magnetic field $\BB_0$ for the Alfv\'{e}n wave (dotted lines),
the slow magnetosonic wave (dashed lines), and the fast magnetosonic wave (solid lines)
for different values of $v_\mathrm{A}^2/c_\mathrm{s}^2$ (shown in legends)
and different values of $v_\mu^2/c_\mathrm{s}^2 =$ 0 (black lines), 0.1 (blue lines),
and 1 (green lines).
}
  \label{fig:Velocity_angle}
\end{figure}

In a compressible fluid, Equations~(\ref{W3})--(\ref{W5}) describe MHD waves
determined by the following dispersion equation:
\begin{eqnarray}
&& \left(\omega^2 - \oA^2\right) \, \Big[\omega^4 - \omega^2 \, ({\bm v}_A^2 + c_{\rm s}^2) k^2 + \oA^2  c_{\rm s}^2 k^2 \Big]
\nonumber\\
&&\quad \quad \quad - \omega^2 \,  (v_\mu \, k)^2 \, \left(\omega^2 - c_{\rm s}^2 k^2\right)=0 ,
\label{W2}
\end{eqnarray}
where $\omega$ is the wave frequency and we have omitted for simplicity the damping effects caused by the magnetic diffusion and kinematic viscosity.
The expression $\left(\omega^2 - \oA^2\right)$ characterizes the Alfv\'{e}n waves,
and the expression in square brackets determines the coupled fast and slow magnetosonic waves.
The last term in Equation~(\ref{W2}) represents the contribution caused by
the CME.
The ratio $\omega/k$ versus the angle $\phi$
between the wavevector ${\bm k}$ and the equilibrium magnetic field $\BB_0$
[obtained by the numerical solutions of Equation~(\ref{W2})] is
shown in Figure~\ref{fig:Velocity_angle}, for different values of
$v_\mu^2/c_\mathrm{s}^2$ and $v_\mathrm{A}^2/c_\mathrm{s}^2$.
Figure~\ref{fig:Velocity_angle} demonstrates that the Alfv\'{e}n wave
and the fast and slow magnetosonic waves in a compressible flow are strongly modified by the CME.
In particular, the CME leads to an increase of the frequencies of the
Alfv\'{e}n wave and of the fast magnetosonic wave and to a decrease of
the frequency of the slow magnetosonic wave.

\section{Laminar dynamos}
\label{sec:laminar-dynamos}

First, we study a kinematic problem concerning the  evolution of the magnetic field
in a given velocity field.
In this problem we neglect the back-reaction of the magnetic field on the velocity field.
We seek a solution of Equation~(\ref{ind-DNS}) for perturbations of the following form:
$\BB(t,x,z)=B_y(t,x,z) {\bm e}_y + \nab   \times  [A(t,x,z) {\bm e}_y]$,
where ${\bm e}_y$ is the unit vector in the $y$-direction.

\subsection{Laminar \protect$v_\mu^2$ dynamo}

We consider the equilibrium configuration:
$\mu_{\rm eq}\equiv \mu_0=$ const and ${\UU}_{\rm eq}=0$.
The functions $B_y(t,x,z)$ and $A(t,x,z)$ are determined by the equations
\begin{eqnarray}
 \frac{\partial A(t,x,z)}{\partial t} &=& v_\mu \, B_y + \eta \Delta A,
\label{A-eq} \\
\frac{\partial B_y(t,x,z)}{\partial t}&=& -v_\mu \, \Delta A + \eta \Delta B_y ,
\label{By-eq}
\end{eqnarray}
where $v_\mu=\eta \, \mu_{0}$,  $\Delta=\nabla_x^2 + \nabla_z^2$,
and the remaining components of the magnetic field are given by $B_x=-\nabla_z A$ and $B_z=\nabla_x A$.
We seek a solution of Equations~(\ref{A-eq})
and~(\ref{By-eq}) of the form $A, B_y \propto \exp[\gamma t + i (k_x x + k_z z)]$.
The growth rate of the dynamo instability is given by
\begin{eqnarray}
\gamma = |v_\mu \, k| - \eta k^2 ,
\label{gamma}
\end{eqnarray}
where $k^2=k_x^2 + k_z^2$.
The dynamo instability is excited $(\gamma> 0)$
for $k < |\mu_{0}|$.
The components of the magnetic field are
\begin{eqnarray}
&& B_x={\rm sgn} \left(\mu_{0}\right) \, {k_z \over k} B_0 \exp(\gamma t) \sin(k_x x + k_z z) ,
\label{F1}\\
&& B_y=B_0 \exp(\gamma t) \cos(k_x x + k_z z),
\label{F2}\\
&& B_z=-{\rm sgn} \left(\mu_{0}\right) \, {k_x \over k} B_0 \exp(\gamma t)
\sin(k_x x + k_z z) .
\label{F3}
\end{eqnarray}
The maximum growth rate of the dynamo instability, attained at
$k \equiv k^{\rm max}$, is given by
\begin{eqnarray}
\gamma^{\rm max} = v_\mu^2 / 4 \eta ,
\label{gamma-max}
\end{eqnarray}
where
\begin{equation}
  \label{eq:87}
k^\max =\frac12|\mu_{0}| .
\end{equation}

\subsection{Laminar $v_\mu^2$--shear dynamo}

\begin{figure}[!t]
\centering
\includegraphics[width=0.5\textwidth]{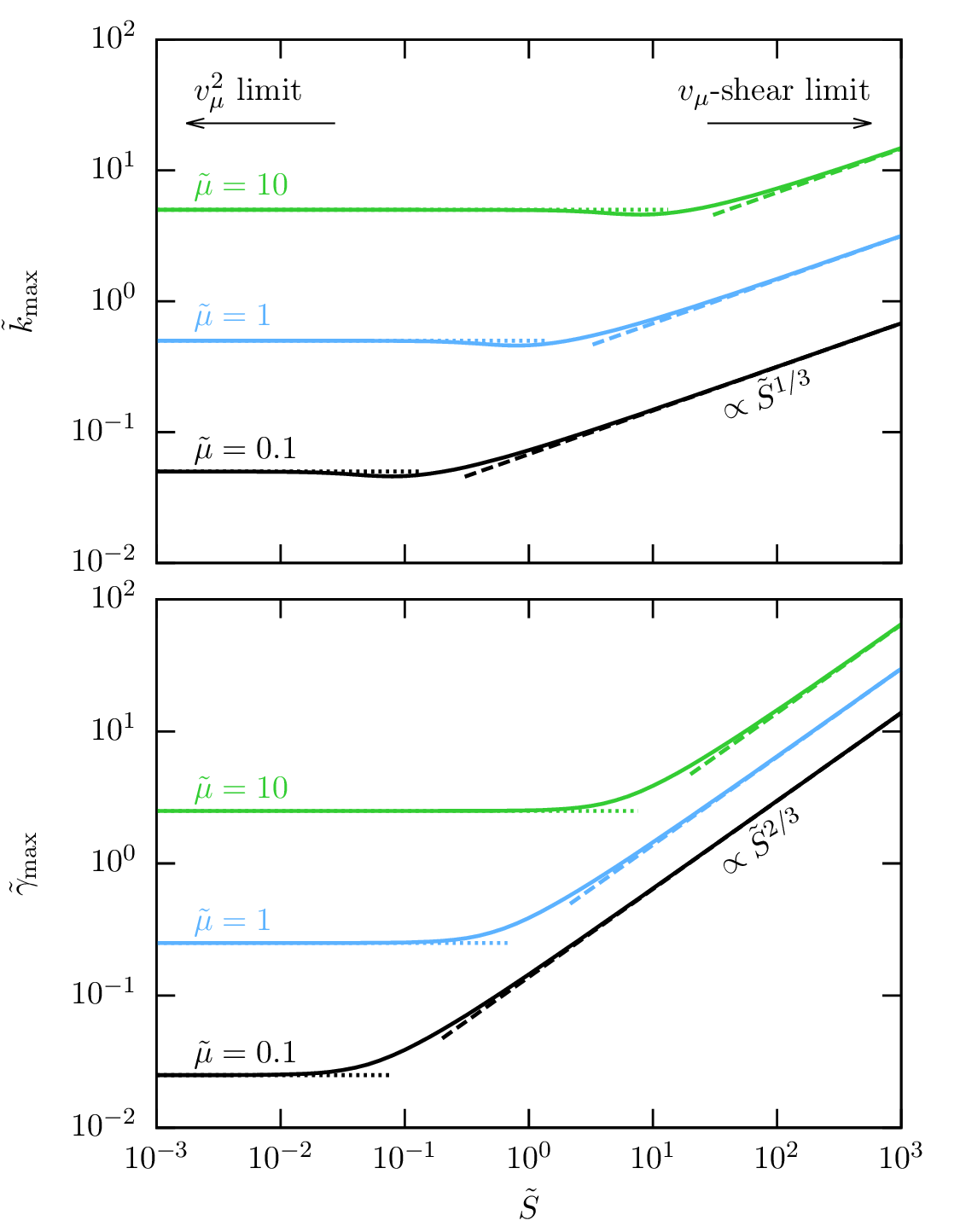}
\caption{
Maximum nondimensional growth rate $\tilde \gamma_\mathrm{max} = \gamma_\mathrm{max}/|v_\mu \, k_\ast|$
(bottom panel) and the nondimensional wavenumber $\tilde k_\mathrm{max} = k_\mathrm{max}/k_\ast$ (top panel)
as a function of the nondimensional shear rate $\tilde S= S/|v_\mu \, k_\ast|$ for different
values of $\tilde \mu = \mu_0 / k_\ast=$ 0.1 (black), 1 (blue), and 10 (green),
where $\gamma_\mathrm{max}=\gamma(k=k_\mathrm{max})$ and $k_x=0$.
Here the wavenumber $k_\ast$ is based on the characteristic scale
of magnetic field variations.
The dotted lines at low $S$ indicate the corresponding values for the
$v_\mu^2$ dynamo [see Equations~(\ref{eq:87}) and~(\ref{gamma-max})],
while the dashed lines at high $S$ indicate the values for the
$v_\mu$--shear dynamo [see Equations~(\ref{kz-max}) and~(\ref{gam-max})].}
\label{fig:max_S}
\end{figure}

Here we consider the equilibrium configuration specified by the shear velocity ${\UU}_{\rm eq}=(0,S\, x,0)$, and $\mu_{\rm eq}\equiv \mu_0=$ const.
This implies that the fluid has a nonzero vorticity
${\bm W} = (0,0,S)$ similar to a differential (nonuniform) rotation.
The functions $B_y(t,x,z)$ and $A(t,x,z)$ are determined by the equations
\begin{eqnarray}
&&\frac{\partial A(t,x,z)}{\partial t} = v_\mu \,  B_y + \eta \Delta A,
\label{A1-eq}\\
&&\frac{\partial B_y(t,x,z)}{\partial t} = - S\nabla_z A - v_\mu \,  \Delta A
+ \eta \Delta B_y .
\label{By1-eq}
\end{eqnarray}
The first term on the rhs of Equation~(\ref{By1-eq})
marks the only difference between the systems of Equations~(\ref{A1-eq})
and (\ref{By1-eq}) and Equations~(\ref{A-eq}) and (\ref{By-eq}).
We look for a solution of Equations~(\ref{A1-eq})
and~(\ref{By1-eq}) of the form $A, B_y \propto \exp[\gamma t + i (k_x x + k_z z- \omega t)]$.
The growth rate of the dynamo instability and the frequency of the dynamo waves are
\begin{eqnarray}
&& \gamma = {|v_\mu \, k| \over \sqrt{2}} \,
\left\{1 + \left[1 + \left({S k_z \over v_\mu \, k^2}\right)^2 \right]^{1\over 2} \right\}^{1\over 2} - \eta k^2 ,
\label{gamma10}\\
&& \omega=  {\rm sgn} \left(\mu_0 k_z\right) \,   {S k_z \over \sqrt{2} k} \,
\left\{1 + \left[1 + \left({S k_z \over v_\mu \,  k^2}\right)^2 \right]^{1\over 2} \right\}^{-{1\over 2}} .
\nonumber\\
\label{omega10}
\end{eqnarray}
This solution describes the laminar $v_\mu^2$--shear dynamo
for arbitrary values of the shear rate.
Using Equation~(\ref{gamma10}) for $k_x=0$, we
determine numerically the maximum growth rate of the dynamo instability,
$\tilde \gamma_\mathrm{max} = \gamma_\mathrm{max}/|v_\mu \, k_\ast|$
(attained at $k \equiv k^{\rm max}$)
versus the shear rate (see Figure~\ref{fig:max_S}).
The laminar $v_\mu^2$--shear dynamo has two limits, one at a low shear rate
(corresponding to the $v_\mu^2$ dynamo; see the dotted lines in
Figure~\ref{fig:max_S}), and another one at a high shear rate
(corresponding to the $v_\mu$--shear dynamo, which will be discussed in the next
section; see the dashed lines in Figure~\ref{fig:max_S}).

\subsection{Laminar $v_\mu$--shear dynamo}

In this section we are interested in a situation where the shear term on the
rhs of Equation~(\ref{By1-eq}) dominates,
i.e., where $|S \nabla_z A| \gg |v_\mu \, \Delta A|$.
This condition implies that
  \begin{equation}
    \label{eq:1}
   1 < {k \over k_z} \ll \frac{S}{v_\mu \, k},
  \end{equation}
i.e., we consider scales with
$k \ll S/ v_\mu$.
The growth rate of the dynamo instability and the frequency of the dynamo waves are
\begin{eqnarray}
&& \gamma = \left({ |v_\mu  \, S \, k_z| \over 2}\right)^{1/2} - \eta k^2 ,
\label{gamma1}\\
&& \omega= {\rm sgn} \left(\mu_{0}  k_z\right) \, \left({|v_\mu  \, S \, k_z| \over 2}\right)^{1/2} ,
\label{omega}
\end{eqnarray}
and we recall that $k^2=k_x^2 + k_z^2$.
The dynamo instability is excited for $k < |v_\mu \, S \, k_z /2\eta^2|^{1/4}$.

The $v_\mu$--shear dynamo mechanism acts as follows:
the nonuniform stretching of the magnetic field component $B_x=-\nabla_z
A$ by the shear or the differential rotation [see the first term on the
right hand side of Equation~(\ref{By1-eq})] causes the generation of a
magnetic field in the $y$-direction.
But the $v_\mu$ effect closes the dynamo loop by generating a magnetic
field in the $x$-direction from the $B_y$-component [see the first term,
$v_\mu B_y$, on the right side of Equation~(\ref{A1-eq})].
The resulting components of the magnetic field are
\begin{eqnarray}
  B_x &=& {\rm sgn} \left(\mu_{0} k_z\right) \,\left|{2 v_\mu \,  k_z \over S}\right|^{1/2} B_0 \exp(\gamma t)
\nonumber\\
&&  \times \sin(k_x x +k_z z - \omega t),
\label{F4}\\
  B_y&=&B_0 \exp(\gamma t) \cos(k_x x +k_z z - \omega t),
\label{F5}\\
 B_z&=& - {\rm sgn} \left(\mu_{0}\right) \, k_x \, \left|{2 v_\mu \over S k_z}\right|^{1/2} B_0 \exp(\gamma t)
\nonumber\\
&&   \times \sin(k_x x +k_z z - \omega t) .
\label{F6}
\end{eqnarray}
The maximum growth rate of the dynamo instability and the maximum frequency
of the dynamo waves, attained at
\begin{equation}
k_z^{\rm max}={1 \over 4} \left({2|S \, v_\mu| \over \eta^2} \right)^{1/3},
\label{kz-max}
\end{equation}
are given by
\begin{eqnarray}
\gamma^{\rm max} &=& {3 \over 8} \left({S^2 \, v_\mu^2 \over 2\eta}\right)^{1/3} - \eta k_x^2,
\label{gam-max}\\
\omega^{\rm max} &=& {{\rm sgn} \left(v_\mu \, k_z\right) \over 2\eta}  \, \left({S^2 \, v_\mu^2 \over 2 \eta}\right)^{1/3} .
\label{omega-max}
\end{eqnarray}

\section{Turbulent large-scale dynamos}
\label{sec:mean-field-dynamo}

In this section, we study large-scale dynamos in
small-scale turbulence with zero mean kinetic helicity.
In the presence of small-scale turbulence,
large-scale properties of the
magnetic field and fluid motion are predicted within the mean-field approach
\citep{M78,P79,KR80,Zeldovich:83},
with all quantities decomposed into mean and fluctuations.
The fluctuating parts have zero mean value: ``overbars'' indicate averaging
over an ensemble of turbulent velocity fields.
We average Equation~(\ref{ind-2}) over the statistics
of the random velocity field:
\begin{eqnarray}
&&\frac{\partial \meanBB}{\partial t} =
\nab   \times   \left(-{\bm \meanEE} + \meanEMF + \meanEMF^\mu \right),
\label{ind-mean}
\end{eqnarray}
where $\meanBB$ is the mean magnetic field.
The total mean electric field, $\meanEE^{\rm tot}$, has three contributions: $\meanEE^{\rm tot}=\meanEE -\meanEMF- \meanEMF^\mu$.
The first contribution is the mean electric field:
\begin{eqnarray}
\meanEE=-\meanUU  \times \meanBB + \eta \, \nab   \times   \meanBB - \meanv_\mu \,
\meanBB ,
\label{mean-E-field}
\end{eqnarray}
where $\meanv_\mu=\eta \, \meanmu$, with $\meanmu$ being the mean chiral chemical potential.
Equation~(\ref{mean-E-field}) is obtained by averaging of Equation~(\ref{eq:94})
for the electric field, $\EE = - \UU\times \BB + \eta \nab\times \BB - v_\mu \, \BB$.
In Equation~(\ref{mean-E-field}) we omitted a small term $-\eta \,
\overline{\mu' \, \bb}$ (which is the second order in $\eta$),
where $\bb$ and $\mu'$ are magnetic fluctuations and chiral chemical
potential fluctuations, respectively.
The second, $-\meanEMF$, and the third, $-\meanEMF^\mu$, contributions to
the total mean electric field, $\meanEE^{\rm tot}$, are related to
the mean electromotive force and discussed in the next section.

\subsection{Mean electromotive force}

The total mean electromotive force is $\bec{\cal E}=\overline{\uu  \times  \bb}
=\meanEMF+\meanEMF^\mu$,
where $\uu$ are the fluctuations of the velocity field,
$\meanEMF$ represents the contributions to the mean electromotive force
in the absence of the CME, and $\meanEMF^\mu$ are the
contributions to the mean electromotive force, caused by the CME.
The general form of the mean electromotive force is given by \citep{R80}:
\begin{eqnarray}
 {\cal E}_{i}(\meanBB) &=& {\alpha}_{ij}(\meanBB) \meanB_{j} - {\eta}_{ij}(\meanBB) (\bec{\nabla} {\bf \times} \meanBB)_{j} + ({\bf V}_{\rm eff}(\meanBB) {\bf \times} \meanBB)_{i}
 \nonumber\\
&&  - [\bec{\delta}(\meanBB) {\bf \times} (\bec{\nabla} {\bf
   \times} \meanBB)]_i - {\kappa}_{ijk}(\meanBB) ({\partial
   \meanB})_{jk} \;,
\label{MB41}
\end{eqnarray}
where $({\partial \meanB})_{ij} = (\nabla_i \meanB_{j} +
\nabla_j \meanB_{i}) / 2$ is the symmetric part of the gradient
tensor of the mean magnetic field $\nabla_i \meanB_{j}$, i.e.,
$\nabla_i \meanB_{j} = (\partial \meanB)_{ij} + \varepsilon_{ijn}
(\bec{\bf \nabla} {\bf \times} \meanBB)_{n} / 2 $,
the last term on the rhs of this expression is
the antisymmetric part of the gradient
tensor of the mean magnetic field,
and $\varepsilon_{ijk}$ is the fully antisymmetric (Levi-Civita) tensor.
Here $\alpha_{ij}(\meanBB)$ and $ \eta_{ij}(\meanBB)$
determine the $\alpha$ effect and turbulent magnetic diffusion,
respectively: ${\bf V}_{\rm eff}(\meanBB)$ is the effective
pumping velocity of the magnetic field: $\kappa_{ijk}(\meanBB)$
describes a contribution to the mean electromotive force related
to the symmetric parts of the gradient tensor of the mean
magnetic field, $({\partial \meanB})_{ij}$, as it appears in
anisotropic turbulence; and finally the $\bec{\delta}(\meanBB)$ term
determines nontrivial behavior of the mean magnetic field in
anisotropic turbulence.
In Equation~(\ref{MB41}) we are neglecting terms $\sim\mathcal{O}(\nabla^2 \meanB_{k})$.

The nonlinear transport coefficients defining the mean electromotive force
and not related to the CME
have been derived in many papers~\citep{RK00,RK01,KR03,RKR03,RK04,RKKB11}.
For nonhelical, isotropic, and inhomogeneous turbulence,
the mean electromotive force $\meanEMF$ in the absence of the CME
is given by
\begin{eqnarray}
\meanEMF = - {\eta}_{_{T}} \, \bec{\nabla} {\bf \times} \meanBB + {\bf V}_{\rm eff} {\bf \times} \meanBB ,
\label{MB120}
\end{eqnarray}
where $\eta_{_{T}} = \ell_0 u_0 / 3$ is the coefficient of turbulent magnetic diffusion
and ${\bf V}_{\rm eff} = - (1/2) \bec{\nabla} {\eta}_{_{T}}$.
In this paper we neglect the effect of large-scale shear on the
mean electromotive force \citep{RK03,RK04,KR08,SS14}.

To derive equations for the nonlinear
coefficients defining the mean electromotive force, we use
equations for fluctuations of velocity, ${\bm u}$, and magnetic fields, ${\bm b}$,
and chiral chemical potential, $\mu'$:
\begin{eqnarray}
 {\partial {\bm u} \over \partial t} &=& {1 \over \meanrho} \Big[- \bec{\nabla}
p_{\rm tot} +  ({\bm b} \cdot\bec{\nabla}) \meanBB + (\meanBB \cdot \bec{\nabla}){\bm
b}\Big] + {\bm F}
\nonumber\\
&&+ \nu \Delta {\bm u} + {\bm u}^N  ,
\label{MB1} \\
{\partial {\bm b} \over \partial t} &=& (\meanBB
\cdot \bec{\nabla}){\bm u} - ({\bm u} \cdot \bec{\nabla}) \meanBB
+ \eta \, \nab   \times   \Big[\meanmu \, {\bb} + \mu' \, \meanBB\Big]
\nonumber\\
&&+ \eta \Delta {\bm b} + {\bm b}^N ,
\label{MB2}\\
\frac{\partial \mu'}{\partial t} &=& -(\uu {\bm \cdot} \nab ) \meanmu + \lambda \,
\eta \, \Big[\bb {\bm \cdot} (\nab   \times   \meanBB) + \meanBB {\bm \cdot} (\nab \times   \bb)
\nonumber\\
&&- \mu' \, {\meanBB}^{\, 2} - 2 \meanmu \, \bb {\bm \cdot} \meanBB \Big]
+ D_5 \Delta \mu' + \mu^N ,
\label{MB3}
\end{eqnarray}
where $\meanrho$ is the mean fluid density; $\meanrho \, {\bm F}$ is a
random external stirring force; ${\bm u}^{N}$, ${\bm b}^{N}$ and $\mu^N$
are the nonlinear terms;
$ p_{\rm tot} = p' + \,(\meanBB \cdot {\bm b}) $ are the fluctuations of total pressure;
and $p'$ are the fluctuations of fluid pressure.
The velocity ${\bm u}$ satisfies to the continuity equation,
$\bec{\nabla} \cdot {\bm u} = 0$, and we consider the case with vanishing mean fluid velocity.

The procedures described in Appendix~\ref{sec:mean-emf} yield
the contribution to the mean electromotive force caused by
the CME for an arbitrary mean magnetic field:
\begin{eqnarray}
\meanEMF^\mu={\bm \alpha}^\mu(\meanBB) \meanB + {\bf V}^\mu_{\rm eff}(\meanBB)
{\bf \times} \meanBB .
\label{MB160}
\end{eqnarray}
The chiral tensor ${\bm \alpha}^\mu(\meanBB) \equiv {\alpha}^\mu_{ij}(\meanBB)$
and the chiral effective pumping velocity ${\bm V}_{\rm eff}^\mu(\meanBB)$ are given
in the next section.

\subsection{The $\alpha_\mu$ effect for a
uniform chiral chemical potential}
\label{AlphamuEffect}

In this section we discuss the $\alpha_\mu$ effect in homogeneous
isotropic incompressible and nonhelical turbulence with a uniform chiral chemical potential.

\subsubsection{Physics of the $\alpha_\mu$ effect}

The mechanism of the $\alpha_\mu$ effect is related to an interaction
between the tangling magnetic fluctuations and the chiral magnetic fluctuations.
To understand the physics of the $\alpha_\mu$ effect,
we discuss here only terms in the induction equation~(\ref{MB2})
that contribute to this effect:
\begin{eqnarray}
{\partial {\bm b} \over \partial t} &=& (\meanBB \cdot \bec{\nabla}){\bm u}
+ \meanv_\mu \nab \times {\bb} + ... ,
\label{MB300}
\end{eqnarray}
where dots denote all other terms in the induction equation~(\ref{MB2})
that contribute to the turbulent diffusion and the chiral effective pumping velocity
(see Appendix~\ref{sec:mean-emf}).
The first term, $(\meanBB \cdot \bec{\nabla}){\bm u}$, on the rhs
of Equation~(\ref{MB300}) describes the production of
the tangling magnetic fluctuations caused by the tangling of the mean magnetic field
$\meanBB$ by sheared small-scale velocity fluctuations.
The second term, $\meanv_\mu \nab \times {\bb}$, in
Equation~(\ref{MB300}) describes the production of
the chiral magnetic fluctuations caused by the interaction
of the fluctuations of the electric current $\nab \times {\bb}$ of the tangling
magnetic fluctuations and the mean chiral chemical potential, $\meanmu$.

Using dimensional analysis, we estimate the level of the tangling magnetic fluctuations
${\bb}_{\rm tang} = \tau \, (\meanBB \cdot \bec{\nabla}){\bm u}$,
and the level of the chiral magnetic fluctuations ${\bb}_{\mu}$:
\begin{eqnarray}
{\bb}_{\mu}= \tau \, \meanv_\mu \nab \times {\bb}_{\rm tang}
= \tau^2 \, \meanv_\mu \nab \times \, [(\meanBB \cdot \bec{\nabla}){\bm u}],
\nonumber\\
\label{MB301}
\end{eqnarray}
where $\tau$ is the characteristic timescale of the random velocity field
to be discussed below.
The mean electromotive force, $\meanEMF^\mu \equiv \overline{\uu \times  {\bb}_{\mu}}$,
caused by the CME is given by
\begin{eqnarray}
\overline{\cal E}_i^\mu = \left(\tau^2 \, \meanv_\mu \,
\overline{u_n \nabla_i \nabla_j u_n} \, \right) \,
\meanB_j \equiv \alpha_{ij}^\mu \, \meanB_j .
\label{MB302}
\end{eqnarray}
Here we took into account that an additional term
in $\overline{\cal E}_i^\mu$ that is proportional
to $\nabla_p \meanB_j$, vanishes in homogeneous and
nonhelical turbulence.
It follows from Equation~(\ref{MB302}) that the chiral $\alpha_{ij}^\mu$ tensor is
\begin{eqnarray}
\alpha_{ij}^\mu = \meanv_\mu \, \tau^2 \, \overline{u_n \nabla_i \nabla_j u_n} =
- \meanv_\mu \, \int \tau^2(k) \, k_i k_j \, \langle{\bm u}^2\rangle_{\bm k}
\, d{\bm k} ,
\nonumber\\
\label{MB303}
\end{eqnarray}
where $\langle{\bm u}^2\rangle_{\bm k} = \overline{{\bm u}^2} \, \tilde E(k) /4 \pi k^2$,
the spectrum function of a random velocity field is
$\tilde E(k)= (q-1) k_0^{-1} (k /k_{0})^{-q}$,
the wavenumber varies in the interval $k_0 < k < k_d$, the exponent
of the spectrum $q$ changes in the interval $1<q<3$,
the wavenumbers $k_{0} = 1 / \ell_{0}$ and $k_d = 1 / \ell_{d}$ with $k_0 \ll k_d$, and $\ell_{d}$ is the dissipation scale.

For small magnetic Reynolds numbers $(\Rm=u_0 \ell_0 /\eta \ll 1)$,
the characteristic timescale of the random velocity field, $\tau$,
is determined by the magnetic diffusion time:
$\tau(k) = 1 / \eta k^2$.
Note that small magnetic Reynolds numbers imply
that $u_0 \ell_0 \ll \eta$.
On the other hand, in Equation~(\ref{eq:94}) for the electric field
we neglected the terms in the second order in $\eta$ which implies that
$\eta \ll c/(4 \pi \, |\nab\Theta|)$.
Combining these two conditions, we obtain that for
small magnetic Reynolds numbers $u_0 \ell_0 \ll \eta \ll c/(4 \pi \, |\nab\Theta|)$.
Substituting the magnetic diffusion timescale into Equation~(\ref{MB303})
and integrating in ${\bm k}$ space, we arrive at the following expression for
the $\alpha_\mu$ effect for $\Rm \ll 1$:
\begin{eqnarray}
\alpha_\mu  = - {(q-1) \over 3(q+1)} \, \Rm^2 \, \meanv_\mu .
\label{MB304}
\end{eqnarray}

For large magnetic Reynolds numbers $(\Rm \gg 1)$, the characteristic time, $\tau$,
of the random velocity field is determined by the turbulent time:
\begin{eqnarray}
\tau(k) = 2 \tau_{0} (k /k_{0})^{1-q} .
\label{MB305}
\end{eqnarray}
For very large fluid Reynolds numbers, ${\rm Re}$, the exponent of the
energy spectrum of turbulent velocity field $q=5/3$ (the Kolmogorov spectrum).
For magnetic Prandtl numbers, $\Pm \equiv \nu/\eta
\geq 1$, the dissipation wavenumber $k_d$ is determined by the Kolmogorov scale,
i.e., $k_d=k_0 {\rm Re}^{3/4}$. Substituting the turbulent timescale~(\ref{MB305})
into Equation~(\ref{MB303}),
and integrating in ${\bm k}$, space we obtain the following expression for
the $\alpha_\mu$ effect for $\Rm \gg 1$ and $\Pm \geq 1$:
\begin{eqnarray}
\alpha_\mu = - {2 \over 3} \meanv_\mu \ln {\rm Re} .
\label{MB306}
\end{eqnarray}
For magnetic Prandtl numbers, $\Pm < 1$, the dissipation wavenumber $k_d$ is
determined by the resistive scale, i.e., $k_d=k_0 \Rm^{3/4}$.
In this case after integration in ${\bm k}$ space we get the expression for
the $\alpha_\mu$ effect for $\Rm \gg 1$ and $\Pm < 1$:
\begin{eqnarray}
\alpha_\mu = - {2 \over 3} \meanv_\mu \ln \Rm .
\label{MB307}
\end{eqnarray}
In the next sections we derive equations for the $\alpha_\mu$ effect using
rigorous approaches.

\subsubsection{Quasi-linear approach}

We consider a kinematic problem of the evolution
of a magnetic field in a given random velocity field.
We start with a weakly nonlinear case in which the nonlinear term in
the induction equation~(\ref{MB2}) is much smaller than the magnetic diffusion term.
This allows us to use the quasi-linear approach.
In the frame of this approach we neglect the nonlinear term in Equation~(\ref{MB2})
but keep the diffusion term.
This implies that the quasi-linear approach is only valid for small magnetic Reynolds numbers.

Next, we apply a multiscale approach.
In the frame of this approach we use the fast
and slow variables, and this allows us to separate small-scale effects (fluctuations)
and large-scale effects (mean fields).
We assume that the maximum scale of random motions $\ell_0$ is much
smaller than the characteristic scales of the spatial variations
of the mean fields, i.e., there is a separation of scales.
Using Equation~(\ref{MB2}) written in Fourier space,
we derive an equation for the cross-helicity tensor
$g_{ij}=\overline{b_i(\omega, {\bm k}) u_j(-\omega, -{\bm k})}$:
\begin{eqnarray}
g_{ij} = G_\eta \, \hat D_{il}^{-1} \, \Big[i({\bm k} {\bf \cdot} \meanBB) \delta_{lm}
- \meanB_{l,m}\Big] f_{mj} + O(\eta^2),
\nonumber\\
\label{MJ14}
\end{eqnarray}
where $f_{ij}=\overline{u_i(\omega, {\bm k}) u_j(-\omega, -{\bm k})}$, the operator
$\hat D_{ij}^{-1} = \delta_{ij} + \tilde \phi \, \varepsilon_{ijm} \, (k_m/k)
+ O(\tilde \phi^2)$ is the inverse of $\hat D_{ij} = \delta_{ij} - \tilde \phi \, \varepsilon_{ijm} k_m$, $\, \delta_{ij}$ is the Kronecker unit tensor,
$\, \tilde \phi = - i k \, G_\eta \, \meanv_\mu$, $\meanB_{i,j} = \nabla_j \meanB_{i}$,
and $G_\eta(k, \omega) = (\eta k^2 + i \omega)^{-1}$.
In Equation~(\ref{MJ14}) we neglected terms $\sim O[\meanB^2; (\nabla \meanBB)^2;
\nabla^2 \meanBB]$.
This method allows us to determine the contribution to the
mean electromotive force $\meanEMF^\mu = \varepsilon_{mji} \int
g_{ij}^\mu(\omega,{\bm k}) \,d {\bm k} \, d\omega$, caused by the
CME: $\meanEMF^\mu = \alpha_\mu  \meanBB$, where the
$\alpha_\mu$ effect is
\begin{eqnarray}
\alpha_\mu  = - {(q-1) \over 3(q+1)} \, \Rm^2 \, \meanv_\mu .
\label{J19}
\end{eqnarray}
Here we took into account that $g_{ij}^\mu(\omega,{\bm k})=- \meanv_\mu \, G_\eta^2 k_i k_j f^{(0)}_{mm}(\omega,{\bm k}) \meanB_j$ (see Equation~(\ref{MJ14})), and
the correlation function $f^{(0)}_{ij}$ with the superscript $(0)$ corresponds
to the background homogeneous isotropic and nonhelical turbulence
with a zero mean magnetic field.
The details of the derivation of the expression for the $\alpha_\mu$ effect
are given in Appendix~\ref{sec:quasi-linear approach}.
Note that Equation~(\ref{J19}) derived using the quasi-linear approach
coincides with Equation~(\ref{MB304}) obtained using the dimensional analysis.

\subsubsection{The $\tau$ approach}

\begin{figure}[!t]
  \centering
  \includegraphics[width=0.45\textwidth]{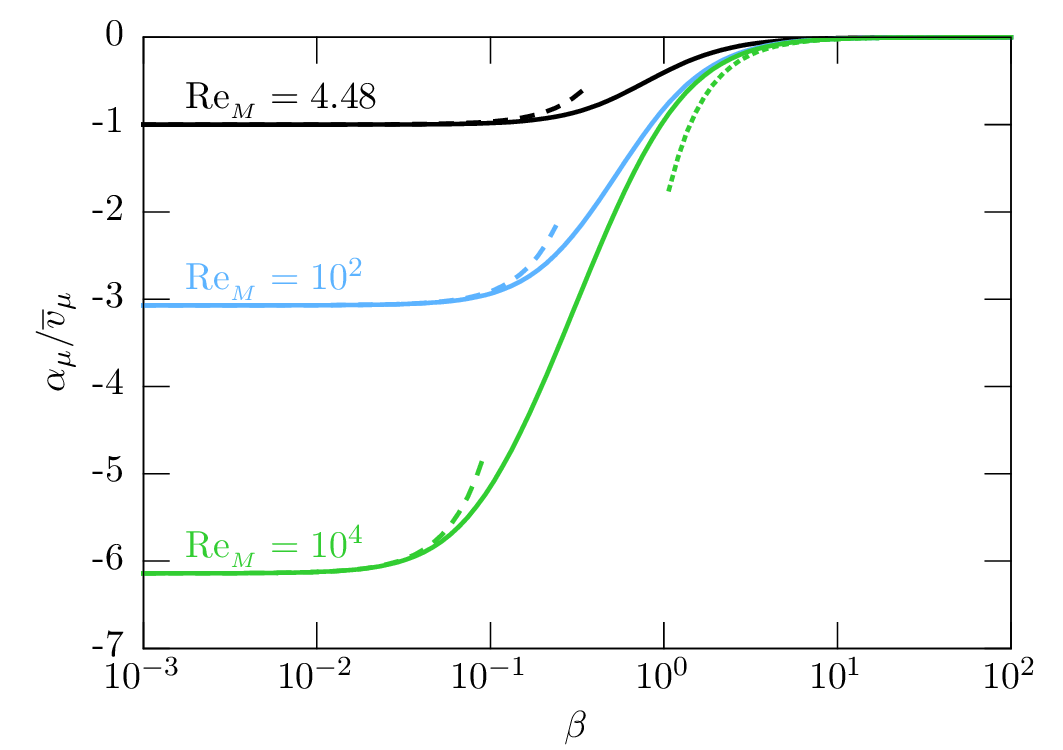}
  \caption{The $\alpha_\mu$ effect as
a function of $\beta=\sqrt{8} \, \, \meanB/B_{\rm eq}$ with
$B_{\rm eq}= \left(\rho \, \overline{{\bm u}^2}\right)^{1/2}$. Solid lines
represent the expression given by Equation~(\ref{MB130B}) normalized by
$\meanv_\mu$, while the dashed lines show the asymptotics at low $\beta$ as given by
(\ref{MB131B}) and at high $\beta$ as given by Equation~(\ref{MB132A}).
The black lines correspond to a fluid Reynolds number of $\mathrm{Re}=4.48$, the blue lines to
$\mathrm{Re}=10^2$, and the green lines to $\mathrm{Re}=10^4$.}
  \label{fig:alpha_beta}
\end{figure}

For large fluid and magnetic Reynolds numbers we use the $\tau$ approach,
which allows us to derive an equation for the contributions to the mean
electromotive force caused by a uniform chiral chemical potential:
$\meanEMF^\mu=\alpha_\mu \meanB$,
where the $\alpha_\mu$ effect is determined by the following expression:
\begin{eqnarray}
&& \alpha_\mu = {4 \over 3} \meanv_\mu \biggl[\ln \biggl({1 + 2 \beta^2 {\rm Re}^{1/2} \over (1 + 2 \beta^2) {\rm Re}^{1/2}}
\biggr) + {1 \over \beta^{2}}\biggl({\arctan (\sqrt{2} \beta) \over \sqrt{2} \beta}
\nonumber\\
&& \quad -1 \biggr) - {1 \over \beta^{2} {\rm Re}^{1/2}}\biggl({\arctan (\sqrt{2} \beta {\rm Re}^{1/4}) \over \sqrt{2} \beta {\rm Re}^{1/4}} -1 \biggr) \biggr] ,
\label{MB130A}
\end{eqnarray}
where $\beta=\sqrt{8} \, \, \meanB/B_{\rm eq}$ and $B_{\rm eq}= (\rho \,
\overline{{\bm u}^2})^{1/2}$ are
the values of the  mean magnetic field under the condition of \emph{equipartition} -- equal fractions of kinetic and magnetic energies.
The details of the derivation of Equation~(\ref{MB130A}) are given in Appendix~\ref{sec:tau approach}.
When $\beta \ll {\rm Re}^{-1/4} \ll 1$, i.e., for a very weak mean magnetic field,
the chiral $\alpha_\mu$ effect is given by
\begin{eqnarray}
\alpha_\mu = - {2 \over 3} \meanv_\mu \ln {\rm Re} \left[1 - {12 \beta^2 {\rm Re}^{1/2} \over 5 \ln {\rm Re}} \right] .
\label{MB131A}
\end{eqnarray}
When ${\rm Re}^{-1/4} \ll \beta \ll 1$, i.e., for a weak mean magnetic field,
the chiral $\alpha_\mu$ effect is
\begin{eqnarray}
\alpha_\mu = - {4 \over 3} \meanv_\mu |\ln (2 \beta^2)| \left[1 + {2 \over 3 |\ln (2 \beta^2)|} \right] ,
\label{MB131AA}
\end{eqnarray}
whereas for $\beta \gg 1$, i.e., for a stronger mean magnetic field,
the chiral $\alpha_\mu$ effect is
\begin{eqnarray}
\alpha_\mu = - {2  \over \beta^2} \meanv_\mu .
\label{MB132A}
\end{eqnarray}
Equations~(\ref{MB130A})--(\ref{MB132A}) are derived for magnetic Prandtl numbers $\Pm
\equiv \nu/\eta \geq 1$.
In the case of $\Pm < 1$, the fluid Reynolds number, ${\rm Re}$, in these equations
should be replaced by the magnetic Reynolds number, $\Rm$.
In this case the $\alpha_\mu$ effect is determined by the following expression:
\begin{eqnarray}
&& \alpha_\mu = {4 \over 3} \meanv_\mu \biggl[\ln \biggl({1 + 2 \beta^2 \Rm^{1/2} \over (1 + 2 \beta^2) \Rm^{1/2}}
\biggr) + {1 \over \beta^{2}}\biggl({\arctan (\sqrt{2} \beta) \over \sqrt{2} \beta}
\nonumber\\
&& \quad -1 \biggr) - {1 \over \beta^{2} \Rm^{1/2}}\biggl({\arctan (\sqrt{2} \beta \Rm^{1/4}) \over \sqrt{2} \beta \Rm^{1/4}} -1 \biggr) \biggr] .
\label{MB130B}
\end{eqnarray}
For a very weak mean magnetic field, $\beta \ll \Rm^{-1/4} \ll 1$,
the $\alpha_\mu$ effect is given by
\begin{eqnarray}
\alpha_\mu = - {2 \over 3} \meanv_\mu \ln \Rm \left[1 - {12 \beta^2 \Rm^{1/2} \over 5 \ln \Rm} \right] .
\label{MB131B}
\end{eqnarray}
For a weak mean magnetic field, $\Rm^{-1/4} \ll \beta \ll 1$, and
for a stronger mean magnetic field, $\beta \gg 1$, the
$\alpha_\mu$ effect is given by Equations~(\ref{MB131AA})
and~(\ref{MB132A}), respectively.
The normalized $\alpha_\mu$ effect as a function of $\beta$ is presented
in Figure~\ref{fig:alpha_beta} for different magnetic Reynolds numbers.
As follows from this section, the $\alpha_\mu$ effect in a
homogeneous turbulence is always negative, and it is opposite to the
$\meanv_\mu$ effect.

\subsection{$\alpha_\mu$ effect and effective pumping velocity
for nonuniform chiral chemical potential}
\label{inhom-turb}

In this section we discuss the $\alpha_\mu$ effect and effective
pumping velocity in inhomogeneous turbulence with a nonuniform chiral
chemical potential.
Using Equations~(\ref{MB100}), (\ref{MB105}), (\ref{MB40}), and (\ref{MB42})
in Appendix~\ref{sec:mean-emf}, we determine contributions to
the functions
$\alpha_{ij}^\mu(\meanBB)$ and ${\bm V}_{\rm eff}^\mu(\meanBB)$
caused by a nonuniform chiral chemical potential and for arbitrary values of the mean
magnetic field:
\begin{eqnarray}
 \alpha_{ij}^\mu(\meanBB) &=&\alpha_\mu \delta_{ij} + {\eta_{_{T}} \eta \, \tau_0 \over 36} \, S(\beta) \biggl[\left(\nabla_i \meanmu\right)\nabla_j \nonumber \\
                          & &  + \left(\nabla_j \meanmu\right)\nabla_i \biggr] \ln \overline{{\bm u}^2}, \label{MB150}\\
 {\bm V}_{\rm eff}^\mu(\meanBB)&=&- {\eta_{_{T}} \eta \, \tau_0 \over 36} \, S(\beta) \left(\bec{\nabla} \meanmu\right)  \times   \left(\bec{\nabla} \ln \overline{{\bm u}^2}\right),
\label{MB151}
\end{eqnarray}
where the isotropic part of the $\alpha$ effect is the sum of the
contribution~(\ref{MB130A}) for $\Pm \geq 1$ (or Equation~(\ref{MB130B})
for $\Pm < 1$) and that caused by the combined action of a nonuniform
chiral chemical potential and inhomogeneous turbulence:
\begin{eqnarray}
\alpha_\mu = {\eta_{_{T}} \eta \, \tau_0 \over 18} \, S(\beta)
\left(\nabla_p \meanmu\right)\left(\nabla_p \ln \overline{{\bm u}^2}\right) .
\label{MB159}
\end{eqnarray}
Here the function $S(x)$ is given in Equation~(\ref{MB105}) of Appendix~\ref{sec:mean-emf}.
Below, we give expressions for ${\alpha}_{ij}^\mu(\meanBB)$ and ${\bm V}_{\rm eff}^\mu(\meanBB)$,
for weak and strong mean magnetic fields.
For a weak field, $\meanB \ll \meanB_{\rm eq}/3$ (i.e., for $\beta \ll 1$), the functions
${\alpha}_{ij}^\mu(\meanBB)$ and ${\bm V}_{\rm eff}^\mu(\meanBB)$ are given by
\begin{eqnarray}
 {\alpha}_{ij}^\mu(\meanBB) &=& {\eta_{_{T}} \,\eta \, \tau_0 \over 6} \, \biggl(\left(\nabla_i
 \meanmu \right) \nabla_j + \left(\nabla_j \meanmu \right) \nabla_i\biggr)
 \ln \overline{{\bm u}^2}
\nonumber \\
&& + \alpha_\mu \delta_{ij},
\label{MB45a}\\
 {\bm V}_{\rm eff}^\mu(\meanBB) &=&- {\eta_{_{T}} \, \eta \, \tau_0\over 6} \, \left(\bec{\nabla} \meanmu\right)
 \times   \left(\bec{\nabla} \ln \overline{{\bm u}^2}\right) .
\label{MB47a}
\end{eqnarray}
while for a strong field, $\meanB \gg \meanB_{\rm eq}/3$, the functions ${\alpha}_{ij}^\mu(\meanBB)$
and ${\bm V}_{\rm eff}^\mu(\meanBB)$ are
\begin{eqnarray}
 {\alpha}_{ij}^\mu(\meanBB) &=& {\eta_{_{T}} \, \eta \, \tau_0\, \meanB_{\rm eq} \over 22 \meanB} \, \left[\left(\nabla_i \meanmu\right)\nabla_j + \left(\nabla_j \meanmu\right)\nabla_i\right] \ln \overline{{\bm u}^2}
\nonumber\\
&& + \alpha_\mu \delta_{ij},
\label{MB48a}\\
{\bm V}_{\rm eff}^\mu(\meanBB) &=& - {\eta_{_{T}} \, \eta \, \tau_0 \,\meanB_{\rm eq} \over 22
\meanB} \left(\bec{\nabla} \meanmu\right)  \times   \left(\bec{\nabla}
\ln \overline{{\bm u}^2}\right) ,
\nonumber \\
\label{MB50a}
\end{eqnarray}
The contribution to $\alpha_\mu$ caused by the combined action of a nonuniform chiral
chemical potential and inhomogeneous turbulence for a weak field is
\begin{eqnarray}
 \alpha_\mu &=& {\eta_{_{T}} \, \eta \, \tau_0 \over 3} \, \left(\nabla_p \meanmu \right)
 \nabla_p \ln \overline{{\bm u}^2},
\label{MB46a}
\end{eqnarray}
while for a strong field it is
\begin{eqnarray}
  \alpha_\mu &=& {\eta_{_{T}} \, \eta \,  \tau_0 \, \meanB_{\rm eq}\over 11 \, \meanB}  \,
\, \left(\nabla_p \meanmu\right)
\, \nabla_p \ln \overline{{\bm u}^2} .
\label{MB49a}
\end{eqnarray}
Note that the chiral transport coefficients,
${\alpha}_{ij}^\mu(\meanBB)$ and ${\bm V}_{\rm eff}^\mu(\meanBB)$, appearing in
the expression for the mean electromotive force vanish when $\eta \to 0$.

\subsection{Generation of the mean kinetic helicity by the CME}

In this section we discuss how the mean kinetic helicity,
$\chi_{_{K}}=\overline{{\bm u} \cdot ({\bm \nabla} \times {\bm u})}$, can be
generated by the CME in nonhelical turbulence.
Using the Navier-Stokes equation for velocity ${\bm U}$ and the equation for vorticity
${\bm W} = {\bm \nabla} \times {\bm U}$ we derive the evolution equation for
the mean kinetic helicity:
\begin{eqnarray}
{\partial \chi_{_{K}} \over \partial t} &=& -2 \overline{{\bm w} \times ({\bm \nabla} \times {\bm b})} \cdot \meanBB - 2 \overline{{\bm w} \times {\bm b}} \cdot ({\bm \nabla} \times \meanBB) - \varepsilon_\chi
\nonumber\\
&&- {\bm \nabla} \cdot {\bm F}_\chi ,
\label{HB1}
\end{eqnarray}
where ${\bm w} = {\bm \nabla} \times {\bm u}$ are the fluctuations of the vorticity, $\varepsilon_\chi \sim \chi_{_{K}} / \tau_0$ is the rate of the dissipation of the mean kinetic helicity, and ${\bm F}_\chi$ is the flux of the mean kinetic helicity:
\begin{eqnarray}
{\bm F}_\chi &=& \overline{{\bm U} \times \left[({\bm \nabla} \times \BB) \times \BB\right]} + \overline{{\bm W} \cdot \left(p/\rho + {\bm U}/2\right)}
\nonumber\\
&&- \overline{{\bm U} \times \left({\bm U} \times {\bm W} \right)} .
\label{HB2}
\end{eqnarray}
To determine the correlation functions $\overline{{\bm w} \times ({\bm \nabla} \times {\bm b})}$ and $\overline{{\bm w} \times {\bm b}}$, we rewrite these functions in ${\bm k}$ space:
$\left[\,\overline{{\bm w} \times ({\bm \nabla} \times {\bm b})}\,\right]_{\bm k} = - \varepsilon_{ijp} k_{n} k_{p} g_{ij}({\bm k})$ and $\left[\, \overline{{\bm w} \times {\bm b}}\, \right]_{\bm k}=i (\delta_{ij} k_{n} - \delta_{nj} k_{i}) g_{ij}({\bm k})$, where $g_{ij}({\bm k})=\overline{b_i({\bm k}) u_j(-{\bm k})}$. To determine the correlation function $g_{ij}({\bm k})$ we use the $\tau$ approach
(see Appendix~\ref{sec:tau approach}).
After integration in ${\bm k}$ space we obtain the contribution to
these correlation functions caused by the CME as
\begin{eqnarray}
\left[\,\overline{{\bm w} \times {\bm b}}\,\right]_\mu = - {\alpha_\mu \over 4} (\meanBB \times {\bm \nabla}) \overline{{\bm u}^2} ,
\label{HB3}
\end{eqnarray}
and $\left[\,\overline{{\bm w} \times ({\bm \nabla} \times {\bm b})}\,\right]_\mu =0$.
This implies that $\left[\,\overline{{\bm w} \times {\bm b}}\,\right]_\mu$
can be nonvanishing only in inhomogeneous turbulence. The generalization
of this result to the case
of inhomogeneous stratified turbulence, where the density stratification
is determined in the anelastic approximation, ${\bm \nabla} \cdot (\rho {\bm b}) =0$,
is performed by the replacement ${\bm \nabla} \to {\bm \nabla} - 2 {\bm \lambda}$, where
${\bm \lambda} = - {\bm \nabla} \rho / \rho$.
Therefore, the evolution of the mean kinetic helicity generated by the CME
is determined by the following equation:
\begin{eqnarray}
{\partial \chi_{_{K}} \over \partial t} &=& {\alpha_\mu \over 2} \left[({\bm \nabla} \times \meanBB) \times \meanBB\right] \cdot {\bm \nabla} \ln \left(\rho^2 \overline{{\bm u}^2}\right) - \varepsilon_\chi
\nonumber\\
&&- {\bm \nabla} \cdot {\bm F}_\chi .
\label{HB4}
\end{eqnarray}
This equation implies that the generation of the mean kinetic helicity
by the CME in nonhelical turbulence is a nonlinear
effect, i.e., it is quadratic in the mean magnetic field, and it occurs
only in inhomogeneous or density-stratified turbulence.
The corresponding $\alpha$ effect caused by the generated mean kinetic
helicity is much smaller than the $\alpha_\mu$ effect considered in
Section~\ref{AlphamuEffect}.

\subsection{Different kinds of turbulent large-scale dynamos}

In this section we consider turbulent large-scale dynamos in the presence
of uniform and nonuniform chiral chemical potentials.
In the case of the nonuniform chiral chemical potential
the $\alpha_\mu$ effect has additional contributions caused by
combined action of the nonuniform chiral chemical potential
and a small-scale inhomogeneous turbulence.
The mean induction equation is given by
\begin{eqnarray}
\frac{\partial \meanBB}{\partial t} &=&
\nab   \times   \biggl\{\meanUU  \times   \meanBB
+ \meanv_\mu \, \meanBB + {\bm \alpha}^\mu \meanBB
+ {\bf V}^\mu_{\rm eff} {\bf \times} \meanBB
\nonumber\\
&& - (\eta+ \, \eta_{_{T}})\nab   \times   \meanBB\biggr\} .
\label{ind4-eq}
\end{eqnarray}
Using this equation, we study different kinds of turbulent large-scale dynamos.
We seek a solution of Equation~(\ref{ind4-eq}) for perturbations of the form
$\meanBB(t,x,z)=\meanB_y(t,x,z) {\bm e}_y + \nab   \times  [\meanA(t,x,z) {\bm e}_y]$,
where ${\bm e}_y$ is the unit vector directed along the $y$-axis.

\subsubsection{Turbulent large-scale $\alpha_\mu^2$ dynamo}

We consider the following equilibrium state: $\meanmu=\meanmu_{\rm eq}=$ const
and ${\bm \meanUU}_{\rm eq}=0$.
The functions $\meanB_y(t,x,z)$ and $\meanA(t,x,z)$ are determined by the following equations:
\begin{multline}
  \frac{\partial \meanA(t,x,z)}{\partial t}
  =(\meanv_\mu + \alpha_\mu)\, \meanB_y
 + (\eta+ \, \eta_{_{T}}) \, \Delta \meanA,
 \label{me-A-eq}
\end{multline}
\begin{multline}
  \frac{\partial \meanB_y(t,x,z)}{\partial t}
  =-(\meanv_\mu + \alpha_\mu) \, \Delta \meanA
 + (\eta+ \, \eta_{_{T}}) \, \Delta \meanB_y ,
\label{me-By-eq}
\end{multline}
where $\Delta=\nabla_x^2 + \nabla_z^2$, and other components of the
magnetic field are $\meanB_x=-\nabla_z \meanA$ and $\meanB_z=\nabla_x
\meanA$.
Mean-field equations~(\ref{me-A-eq}) and (\ref{me-By-eq})
in the presence of a small-scale turbulence are different from
Equations~(\ref{A1-eq}) and (\ref{By1-eq}) used in Section~\ref{sec:laminar-dynamos}
for studying the laminar dynamo effects. In particular,
these mean-field equations contain two new terms related to
(i) the $\alpha_\mu$ effect (ii) the turbulent diffusion $\eta_{_{T}}$;
and
(iii) the $v_\mu$ effect is replaced
in the mean-field equations by the mean $\meanv_\mu$ effect.
We are working under the assumption that the ratio of $\eta_{_{T}}/ \eta$ related to
the \emph{magnetic Reynolds number,}
\begin{equation}
  \label{eq:88}
  \Rm \equiv \frac{3\eta_{_{T}}}\eta
\end{equation}
is \emph{large}, i.e., $\Rm\gg 1$.
This is the case under for many astrophysical flows, e.g., in the early
universe \citep{Banerjee:04,Banerjee:03,Jedamzik:96}, in stellar and galactic
dynamos \citep{M78,P79,KR80,Zeldovich:83}.

We are looking for a solution of the mean-field equations~(\ref{me-A-eq})
and~(\ref{me-By-eq}) in the form
$$\meanA, \meanB_y \propto \exp[\gamma t + i (k_x x + k_z z)].$$
The growth rate of the dynamo instability is given by
\begin{eqnarray}
\gamma = |(\meanv_\mu + \alpha_\mu)\, k| - (\eta+ \, \eta_{_{T}}) \, k^2 ,
\label{me-gamma}
\end{eqnarray}
and $k^2=k_x^2 + k_z^2$.
The components of the mean magnetic field are
\begin{eqnarray}
  \meanB_x&=&{\rm sgn} \left(\meanv_\mu + \alpha_\mu\right) \, {k_z \over k} \meanB_0
  \exp(\gamma t) \sin(k_x x + k_z z) ,
\nonumber\\
\label{F10}\\
  \meanB_y&=&\meanB_0 \exp(\gamma t) \cos(k_x x + k_z z),
\label{F20}\\
  \meanB_z&=&-{\rm sgn} \left(\meanv_\mu + \alpha_\mu\right) \, {k_x \over k} \, \meanB_0
  \exp(\gamma t) \sin(k_x x + k_z z) .
\nonumber\\
\label{F30}
\end{eqnarray}
The maximum growth rate of the dynamo instability, attained at
$k \equiv k^{\rm max}=|\meanv_\mu + \alpha_\mu|/2(\eta+ \, \eta_{_{T}})$, is given by
\begin{eqnarray}
\gamma^{\rm max} = {(\meanv_\mu + \alpha_\mu)^2\over 4 (\eta+ \, \eta_{_{T}})}
= {(\meanv_\mu + \alpha_\mu)^2\over 4 \eta \, (1 + \, \Rm/3)}.
\label{me-gamma-max}
\end{eqnarray}
For small magnetic Reynolds numbers, this equation
yields the correct result for the laminar $v_\mu^2$ dynamo
[see Equation~(\ref{gamma-max})].
For large magnetic Reynolds number, the maximum growth rate of
the dynamo instability decreases with $\Rm$, i.e.,
$\gamma^{\rm max} \propto \Rm^{-1}$.

Since the $\alpha_\mu$ effect in a homogeneous turbulence is always negative
while the $\meanv_\mu$ effect is positive, the $\alpha_\mu$ effect
decreases the $\meanv_\mu$ effect.
Both effects compensate each others at $\Rm=4.48$ (see Figure~\ref{fig:alpha_beta}).
However, for large fluid and magnetic Reynolds numbers, $\meanv_\mu \ll |\alpha_\mu|$,
so we can neglect $\meanv_\mu$ in the equations of this section.
This case corresponds to the turbulent large-scale $\alpha_\mu^2$ dynamo.

\subsubsection{Turbulent large-scale $\alpha_\mu^2$--shear dynamo}

Let us consider an equilibrium with  mean velocity shear, $\meanS$, i.e.,
$\meanUU_{\rm eq}=(0,\meanS x,0)$, and $\meanmu=\meanmu_{\rm eq}=$ const.
The functions $\meanB_y(t,x,z)$ and $\meanA(t,x,z)$ are determined by the following equations:
\begin{eqnarray}
\frac{\partial \meanA(t,x,z)}{\partial t} &=& (\meanv_\mu + \alpha_\mu) \, \meanB_y
+ (\eta+ \, \eta_{_{T}}) \, \Delta \meanA,
\label{meanA-eq}\\
\frac{\partial \meanB_y(t,x,z)}{\partial t} &=& - \meanS \, \nabla_z \meanA
-(\meanv_\mu + \alpha_\mu) \, \Delta \meanA
\nonumber\\
&& + (\eta+ \, \eta_{_{T}}) \, \Delta \meanB_y .
\label{meanBy-eq}
\end{eqnarray}
We seek a solution of Equations~(\ref{meanA-eq})
and~(\ref{meanBy-eq}) of the form $$\meanA, \meanB_y \propto \exp[\gamma t + i (k_x x + k_z z
- \omega t)].$$
The growth rate of the dynamo instability and the frequency of the dynamo waves are given by
\begin{multline}
 \gamma= {|(\meanv_\mu + \alpha_\mu)  \, k| \over \sqrt{2}} \,
\sqrt{1 + \sqrt{1 + \left({\meanS k_z \over (\meanv_\mu + \alpha_\mu) \, k^2}\right)^2
}}\\
 - (\eta+ \, \eta_{_{T}}) \, k^2 ,
\label{gamma100}
\end{multline}
\begin{multline}
\omega= {\meanS k_z\over \sqrt{2} k} \,
\left\{1 + \left[1 + \left({\meanS k_z \over (\meanv_\mu + \alpha_\mu) \,  k^2}\right)^2
\right]^{1\over 2} \right\}^{-{1\over 2}}
\\\times {\rm sgn} \left[(\meanv_\mu + \alpha_\mu)  k_z\right] .
\label{omega100}
\end{multline}
For small magnetic Reynolds numbers, these equations
yield the correct results for the laminar $v_\mu^2$-shear dynamo;
see Equations~(\ref{gamma10}) and~(\ref{omega10}).
The dependencies of the maximum dimensionless growth rate
$\tilde \gamma_\mathrm{max} = \gamma_\mathrm{max}/|(\meanv_\mu + \alpha_\mu)\, k_\ast|$
and of the dimensionless wavenumber $\tilde k_\mathrm{max} = k_\mathrm{max}/k_\ast$
on the nondimensional shear rate $\tilde S= S/|(\meanv_\mu + \alpha_\mu) \, k_\ast|$
are similar to those shown in Figure~\ref{fig:max_S},
after the change $\tilde \mu \to (\meanv_\mu + \alpha_\mu)/[k\eta(1 + \Rm/3)]$.

In the case of very large fluid and magnetic Reynolds numbers, $\meanv_\mu \ll |\alpha_\mu|$,
so we can neglect $\meanv_\mu$ in the equations of this section.
This case corresponds to the turbulent large-scale $\alpha_\mu^2$--shear dynamo
for an arbitrary value of the shear.

\subsubsection{Turbulent large-scale $\alpha_\mu$--shear dynamo}

Next, we consider a plasma where the shear term
in Equation~(\ref{meanBy-eq}) dominates, i.e., we assume that $k^2 \, |(\meanv_\mu + \alpha_\mu)
/ k_z | \ll |\meanS|$.
The growth rate of the dynamo instability and the frequency of the dynamo waves are
\begin{eqnarray}
&& \gamma = \left({|(\meanv_\mu + \alpha_\mu) \, \meanS \, k_z| \over 2}\right)^{1/2}
- (\eta+ \, \eta_{_{T}}) \, k^2 ,
\label{m-gamma1}\\
&& \omega= {\rm sgn} \left[(\meanv_\mu + \alpha_\mu)\, k_z\right] \, \left({(\meanv_\mu + \alpha_\mu)
\, \meanS \, k_z| \over 2}\right)^{1/2} .
\nonumber\\
\label{m-omega}
\end{eqnarray}
The components of the mean magnetic field are
\begin{eqnarray}
  \meanB_x &=& {\rm sgn} \left[(\meanv_\mu + \alpha_\mu) \,  k_z\right]
  \,\left|{2 (\meanv_\mu + \alpha_\mu) \, k_z
  \over \meanS}\right|^{1/2} \meanB_0 \exp(\gamma t)
\nonumber\\
&&  \times \sin(k_x x +k_z z - \omega t),
\label{F40}\\
  \meanB_y &=& \meanB_0 \exp(\gamma t) \cos(k_x x +k_z z - \omega t),
\label{F50}\\
  \meanB_z&=& - {\rm sgn} \left(\meanv_\mu + \alpha_\mu\right) \, k_x \,
  \left|{2 (\meanv_\mu + \alpha_\mu)
  \over \meanS k_z}\right|^{1/2} \meanB_0 \exp(\gamma t)
\nonumber\\
&&   \times \sin(k_x x +k_z z - \omega t) .
\label{F60}
\end{eqnarray}
The maximum growth rate of the dynamo instability and the maximum frequency
of the dynamo waves, attained at $k_x=0$ and
\begin{eqnarray}
k_z^{\rm max}={1 \over 4} \left({2 |\meanS \, (\meanv_\mu + \alpha_\mu)| \over (\eta
+ \, \eta_{_{T}})^2} \right)^{1/3},
\label{m-kz-max}
\end{eqnarray}
are given by
\begin{eqnarray}
\gamma^{\rm max} &=& {3 \over 8} \left({\meanS^2 \, (\meanv_\mu + \alpha_\mu)^2 \, \over 2 (\eta
+ \, \eta_{_{T}})} \right)^{1/3},
\label{m-gam-max}\\
\omega^{\rm max} &=& {{\rm sgn} \left[(\meanv_\mu + \alpha_\mu) \, k_z\right]
\over 2\eta} \left[{\meanS^2 \, (\meanv_\mu + \alpha_\mu)^2 \, \over 2(\eta+ \, \eta_{_{T}})}
\right]^{1/3} .
\nonumber\\
\label{m-omega-max}
\end{eqnarray}

In the case of the nonuniform chiral chemical potential
we assumed that the generated $\alpha$ effect due to
the combined action of the large-scale shear
and inhomogeneous turbulence is smaller than
the additional contributions to the $\alpha_\mu$ effect caused by
the combined action of the nonuniform chiral chemical potential
and a small-scale inhomogeneous turbulence.
The $\alpha$ effect is estimated by $\alpha=S \ell_0^2/L_u$,
where $L_u$ is the characteristic scale of the inhomogeneity of the turbulence.
The above condition implies that $|\meanS| \ll \eta |\nab  \meanmu|$.
The large-scale shear must satisfy the condition
$|\alpha_\mu k_z | \ll |\meanS|$ (see above).
These two conditions imply that $k_z \ll L_u/\ell_0^2$.
If this bound is not satisfied, the contribution to the CME caused by
the combined action of the nonuniform chiral chemical potential
and a small-scale inhomogeneous turbulence is not important.

\subsection{Dynamic nonlinearity in mean-field dynamos}
\label{sec:cons-laws}

In this section we discuss the dynamic nonlinearity,
which can play an important role in nonlinear large-scale magnetic
dynamos and in the presence of the CME.

\subsubsection{Mean fields}

We average Equation~(\ref{AA-2}) over the random velocity field:
\begin{eqnarray}
&&\frac{\partial \meanAA}{\partial t} = - {\bm \meanEE} + \meanEMF + \meanEMF^\mu
+ \nab  \meanPhi ,
\label{AA-mean}
\end{eqnarray}
where $\meanBB = \nab   \times   \meanAA$.
Multiplying Equation~(\ref{ind-mean}) by $\meanAA$ and Equation~(\ref{AA-mean}) by $\meanBB$, and adding them, we obtain an evolutionary equation for the mean magnetic helicity density, $\meanAA {\bm \cdot} \meanBB$:
\begin{eqnarray}
&&\frac{\partial \meanAA {\bm \cdot} \meanBB}{\partial t} + \nab  {\bm \cdot}
\Big[{\bm \meanEE} \times   \meanAA -(\meanEMF + \meanEMF^\mu) \times   \meanAA
+ \meanBB \, \meanPhi\Big]
\nonumber\\
&& \quad  = - 2 {\bm \meanEE} {\bm \cdot} \meanBB + 2 (\meanEMF + \meanEMF^\mu) {\bm \cdot} \meanBB .
\label{AB-mean}
\end{eqnarray}
Averaging Equation~(\ref{mu-2}) over the random velocity field, we find that
\begin{eqnarray}
&&\frac{\partial (2 \meanmu/\lambda)}{\partial t} + \nab  {\bm \cdot}
\big[- (2D_5/\lambda) \, \nab  \meanmu\big] = 2 {\meanEE} {\bm \cdot} {\meanBB}
-2 \meanEMF^\mu {\bm \cdot} \meanBB
\nonumber\\
&& \quad + 2 \eta \, \Big[\overline{{\bm \bb} \, (\nab   \times   {\bm \bb})}
- \meanmu \, \, \overline{{\bm \bb}^2}\Big] .
\label{mu-mean}
\end{eqnarray}
Adding Equations~(\ref{AB-mean}) and~(\ref{mu-mean}) we obtain an equation for
$\meanAA {\bm \cdot} \meanBB + 2 \meanmu/\lambda$, namely,
\begin{multline}
\frac{\partial}{\partial t} \Big(\meanAA {\bm \cdot} \meanBB
+ 2 \meanmu/\lambda\Big) \\+ \nab  {\bm \cdot}
\Big[{\bm \meanEE} \times   \meanAA -(\meanEMF + \meanEMF^\mu) \times   \meanAA
+ \meanBB \, \meanPhi
 \quad - (2D_5/\lambda) \, \nab  \meanmu  \Big]\\
= 2 \meanEMF {\bm \cdot} \meanBB
+ 2 \eta \, \Big[ \overline{\bb \, (\nab   \times   \bb)}
- \meanmu \, \, \overline{\bb^2} \Big].
\label{mean}
 \end{multline}
Substituting in Equation~(\ref{CL})
${\bm A}=\meanAA + {\bf a}$, $\BB=\meanBB+\bb$, $\EE=\meanEE+\EE '$,
$\mu=\meanmu + \mu '$, $\Phi=\meanPhi + \phi' $ and averaging the equation so obtained
over the random velocity field, we get
\begin{eqnarray}
&&\frac{\partial}{\partial t} \Big(\meanAA {\bm \cdot} \meanBB
+ 2 \meanmu/\lambda + \overline{{\bf a} {\bm \cdot} \bb} \Big) + \nab  {\bm \cdot}
\Big[{\bm \meanEE} \times   \meanAA -(\meanEMF + \meanEMF^\mu) \times   \meanAA
\nonumber\\
&& \quad + \meanBB \, \meanPhi - (2D_5/\lambda) \, \nab  \meanmu
+ \overline{\EE '  \times   {\bf a}} + \overline{\bb \phi'} \Big] = 0 .
\label{CL-mean}
\end{eqnarray}

\subsubsection{Equation for fluctuations of magnetic helicity density}

Subtracting Equation~(\ref{mean}) from Equation~(\ref{CL-mean}), we obtain
an evolution equation for the small-scale magnetic helicity density,
$\chi_m=\overline{{\bf a} {\bm \cdot} \bb}$, namely,
\begin{multline}
\frac{\partial \chi_m}{\partial t} + \nab  {\bm \cdot}
\Big[\overline{\EE '  \times   {\bf a}} + \overline{\bb \phi'} \Big]\\
= - 2 \meanEMF {\bm \cdot} \meanBB
- 2 \eta \, \Big[ \overline{\bb \, (\nab   \times   \bb)}
\quad - \meanmu \, \, \overline{\bb^2}\Big] .
\label{mag-hel-small-scale}
 \end{multline}
This equation, taking into account the CME, plays
a crucial role in the nonlinear stage of the large-scale (mean-field)
dynamo evolution.
Without the CME, it has been derived and used for the
investigation of the nonlinear evolution of the mean magnetic field in
a number of studies \citep{KR82,KRR95,GD94,KR99,KMRS2000,BF00,BB02,BS05}.
The magnetic fluctuations, $\overline{\bb^2}$, are determined in \citet{RK07}.
Equation~(\ref{mag-hel-small-scale}) can be used in mean-field simulations
of the nonlinear large-scale magnetic dynamos in the presence of the CME.

\section{Chiral MHD equations in an expanding universe}
\label{sec:chiral-mhd-equations}

In this section we demonstrate that basic properties of the chiral MHD
equations, analyzed in this paper, also hold in an expanding universe.
There are many excellent reviews where the subject of
ordinary MHD in an expanding universe is discussed \citep[see,
e.g.][]{Barrow:2006ch,Subramanian:09,Subramanian:16,Durrer:13}.
Therefore, we will discuss here only the novelties, brought by the
presence of the axial current and axial anomaly.

\subsection{Axial anomaly in an expanding universe}
\label{sec:axial-anom-FRW}

The axial anomaly in a curved background with metric $g_{\mu\nu}$ has the form
\begin{equation}
  \label{eq:24}
  \pfrac{}{x^\mu}\Bigl(\sqrt{-g} j_5^\mu\Bigr) =
  \frac{\alphaem}{4\pi\hbar}\bar{\epsilon}^{\mu\nu\lambda\rho} F_{\mu\nu} F_{\lambda\rho} ,
\end{equation}
where $g = \mbox{det}(g_{\mu\nu})$ and $\bar{\epsilon}^{\mu\nu\lambda\rho}$
is a flat-space antisymmetric tensor
(e.g., $\bar{\epsilon}^{0123} = +1$).
The expanding universe is described by the metric
\begin{equation}
  \label{eq:14}
  ds^2 = -c^2dt^2 + a^2(t)d\bm x^2
\end{equation}
with $\sqrt{-g} = a^3(t)$.
As discussed in detail by \cite{Subramanian:09,Subramanian:16}, an
observer measures physical quantities in a local inertial frame.
This implies that for the current density $\JJ$, for example,
the corresponding 4-vector is $j^\mu=(\varrho,J^i/a)$.

In order to recast Equation~\eqref{eq:24} in a form similar to that in
flat space, we define \emph{electric} and \emph{magnetic} fields in terms of
the components of the field strength tensor $F_{\mu\nu}$ \citep[see][for
details]{Brandenburg:96,Subramanian:09}:
\begin{equation}
  \label{eq:22}
  F_{0i} = a\, \EE^i,\quad F_{ij} = a^2\,\epsilon_{ijk}\BB^k \;,
\end{equation}
and write
\begin{equation}
  \label{eq:25}
  \bar\epsilon^{\mu\nu\lambda\rho} F_{\mu\nu} F_{\lambda\rho} = 8 a^3(t) \EE \cdot \BB .
\end{equation}
Hence, Equation~\eqref{eq:24} becomes
\begin{equation}
  \label{eq:26}
  \frac1{a^3}\frac{\partial\bigl(a^3 n_5)}{\partial t} + \nab\cdot(\JJ_5/a)  = \frac{2\alphaem}{\pi\hbar}\EE\cdot \BB .
\end{equation}
where $j_5^\mu = (c n_5, \JJ_5/a)$.
By introducing comoving quantities \citep{Brandenburg:96,Banerjee:04,Subramanian:09}:
\begin{equation}
  \label{eq:18}
  \begin{aligned}
    \tilde \EE & \equiv a^2 \EE\\
    \tilde \BB & \equiv a^2 \BB\\
    \tilde n_5 &\equiv a^3 n_5\\
    \tilde \JJ_5 &\equiv {a^3} \JJ_5
  \end{aligned}
\end{equation}
and switching to the conformal time, $\tauc$,
\begin{equation}
d\tauc \equiv \frac{dt}{a(t)},
\label{eq:38}
\end{equation}
we obtain
\begin{equation}
  \label{eq:27}
  \pfrac{\tilde n_5}{\tauc} + \nab \cdot\tilde{\JJ}_5
  = \frac{2\alphaem}{\pi\hbar}
  \tilde{\EE}\cdot \tilde{\BB}
\end{equation}
an expression identical to Eq.~\eqref{eq:5}.

The quantity $n_5$ is related to $\mu_5$ as
$n_5 = \mu_5 (k_BT)^2/(6\hbar^3 c^3)$,
where the temperature $T$ now depends on time.  Therefore, if one introduces
\begin{equation}
  \label{eq:68}
  \tilde \mu_5 \equiv a \mu_5 ,
\end{equation}
the relation between $\tilde n_5$ and $\tilde \mu_5$ is given by
\begin{equation}
  \label{eq:69}
  \tilde n_5 \approx \frac{k_B^2 \bigl(a\, T\bigr)^2}{6\hbar^3 c^3} \tilde\mu_5 .
\end{equation}
Finally, introducing a ``comoving'' axion field,
\begin{equation}
  \label{eq:70}
  \frac{D\tilde \Theta}{D\tauc} \equiv \pfrac{\tilde \Theta}{\tauc}+ \tilde U \cdot \nab \tilde\Theta \equiv \frac{\alphaem}{\pi\hbar} \tilde \mu_5 ,
\end{equation}
and setting $\tilde \Lambda$ via
\begin{equation}
  \label{eq:71}
  \tilde\Lambda^{-2} \equiv \frac{12 \alphaem}\pi\, \frac{\hbar^2 c^3}{k_B^2 (a\,T)^2} ,
\end{equation}
we find that the overall property of relativistic MHD \citep{Brandenburg:96,Banerjee:04,Jedamzik:96} holds for chiral MHD as well.

\section{Conclusions}
\label{sec:discussion}

In this paper we have investigated laminar and turbulent dynamos in chiral MHD.
The chiral effect occurs owing to relativistic fermions in a magnetized plasma
and plays an important role.
It may explain the origin and evolution of magnetic fields in the early
universe, and has applications to the theory of neutron stars and
quark--gluon plasmas.
To study the different dynamo effects, we use the nonlinear system of
chiral MHD equations (\ref{ind-DNS})--(\ref{mu-DNS}), where we take into account the feedback of the magnetic field onto the chiral chemical potential in the hydrodynamic
flow of the plasma.
The sum of magnetic helicity and normalized chiral
chemical potential is strictly conserved -- independent of the magnetic
Reynolds number.
This determines the main nonlinearity in the system.

We have considered the one-fluid MHD plasma approximation and studied
the modifications of MHD waves due to the CME.
We have analyzed three kinds of waves in the plasma, namely, Alfv\'{e}n
waves and fast and slow magnetosonic waves.
We have demonstrated that the CME decreases the
frequency of Alfv\'{e}n waves for an incompressible fluid, increases
the frequencies of Alfv\'{e}n waves and fast magnetosonic waves for a
compressible flow, and decreases the frequency of slow magnetosonic waves.

The CME has been shown to be responsible for new dynamos.
The latter originate from a new term in the chiral induction equation (\ref{ind-DNS}) which is proportional to $v_\mu=\eta \, \mu$, where $\eta$ is the resistive (Ohmic) magnetic diffusion and $\mu$ is the chiral chemical potential.
In a plasma without turbulence, there are laminar $v_\mu^2$ dynamos
(discussed previously) and laminar $v_\mu$--shear dynamos (or laminar $v_\mu^2$--shear dynamos) in sheared fluid flows.
In the $v_\mu^2$ dynamo, all components of the magnetic field are generated by the $v_\mu$ effect.
In the $v_\mu$--shear dynamo, the magnetic field component perpendicular to the shear velocity is stretched by the shear velocity.
This generates a component of the magnetic field along the shear velocity.
The $v_\mu$ effect caused by the CME closes the
dynamo loop by generating components of the magnetic field perpendicular
to the shear velocity.

In turbulent flows with nonzero mean kinetic helicity, the usual
$\alpha$ effect is caused by a mean kinetic helicity, independently of
the resistive (Ohmic) magnetic diffusion.
In such turbulent flows with large fluid and magnetic Reynolds numbers
the CME is not important.

However, in turbulent flows \emph{with zero mean kinetic helicity}, the
CME plays a crucial role and contributes to the mean electromotive force.
For large magnetic and fluid Reynolds numbers,
a new $\alpha_\mu$ effect that is caused by an interaction of
the CME and fluctuations of the small-scale
current produced by tangling magnetic fluctuations is dominant.
The tangling magnetic fluctuations are produced by tangling of
the large-scale magnetic field by sheared velocity fluctuations.
The $\alpha_\mu$ effect
causes turbulent $\alpha_\mu^2$ and $\alpha_\mu$--shear
(or $\alpha_\mu^2$--shear) dynamos.
In turbulent flows with large magnetic Reynolds numbers, the turbulent magnetic diffusion, $\eta_{_{T}}$, is much larger than the resistive (Ohmic) magnetic diffusion $\eta$.
This implies that the turbulent magnetic diffusion increases the characteristic scale of the mean magnetic field.

\begin{acknowledgements}
We thank Dmitri Kharzeev, Michael Shaposhnikov, and Kandaswamy Subramanian
for stimulating discussions.
This project has received funding from the European Research Council
(ERC) under the European Union’s Horizon 2020 research and innovation
program (GA no.\ 694896).
This work has been supported by
the NSF Astrophysics and Astronomy Grant Program (grant 1615100),
the Research Council of Norway
under the FRINATEK (grant no.\ 231444) and by the European Research Council under the NuBSM grant (no.\ 694896).
IR and NK thank NORDITA and LASP (Colorado University) for hospitality
and support during their visits.
IR thanks Ecole Polytechnique F\'{e}d\'{e}rale de Lausanne
for hospitality and support during his visit.
This work has been initiated while participating at
the NORDITA program on
``Origin, Evolution, and Signatures of Cosmological Magnetic Fields''
during 2015 June 15--July 10.
\end{acknowledgements}

\appendix

\section{Current along the magnetic field:
Pedagogical derivation}
\label{sec:CME_current_LLL}

In this appendix we will explain the origin of the current given by Equation~\eqref{eq:13} in the simplest setup.
It has been first derived by~\cite{Vilenkin:80a} using the method close to the one sketched here and independently rederived in a number of different ways
\citep{Redlich:1984md,Tsokos:85,Alekseev:98a,Frohlich:2000en,Frohlich:2002fg,Kharzeev:07,Son:2009tf,Fukushima:08}.
There are many excellent reviews on the subject;
see, e.g., \citet{Kharzeev:12a}.
This appendix is intended just to give a basic idea.

Let us consider a uniform magnetic field $\BB= (0,0,B)$.
The spectrum of the fermions is given by the following expression;
see Section~32 in \cite{Landau-vol4}:
\begin{eqnarray}
  \label{eq:10}
&&  E_{\bm p} = \pm \sqrt{p_z^2 c^2 + \hbar c |e| B(2n+1 - 2 s_z)}
\end{eqnarray}
where $n = 0,1,2\dots$;
$s_z = \pm \frac12$ is the projection of the fermion's spin on the magnetic field ($z$-axis);
$E_{\bm p}$ is the particle energy; and $p_z$ is the $z$ component of particle momentum.
As a result, for $n=0$ (\emph{lowest Landau level})
and for positive
$s_z = +\frac 12$, the motion of particles is that of \emph{free
one-dimensional massless fermions} with $E_{\bm p} = \pm c|\bm p|$.
Thus, the particles with $p_z > 0$ have positive projection
of spin $\bm s$ onto momentum $\bm p$ (\emph{right-chiral} particles),
and the particles with $p_z < 0$ have negative projection of the
spin onto momentum (left-chiral particles).
In vacuum (at zero temperature and zero chemical potential), all the
states with $E_{\bm p} < 0$ are filled (the \emph{Dirac sea}), whereas
all the states with $E_{\bm p} > 0$ are empty.
If an electric field is applied parallel to the magnetic field, one
will observe the disappearance of left-chiral particles and the appearance of
right-chiral anti-particles (or vice versa, depending on the sign of the
electric field), so that the total electric charge is conserved.
This is the manifestation of the axial anomaly; see the discussion in, e.g., \cite{Volovik:98}.

\begin{figure}[!t]
  \centering
  \includegraphics[width=0.45\textwidth]{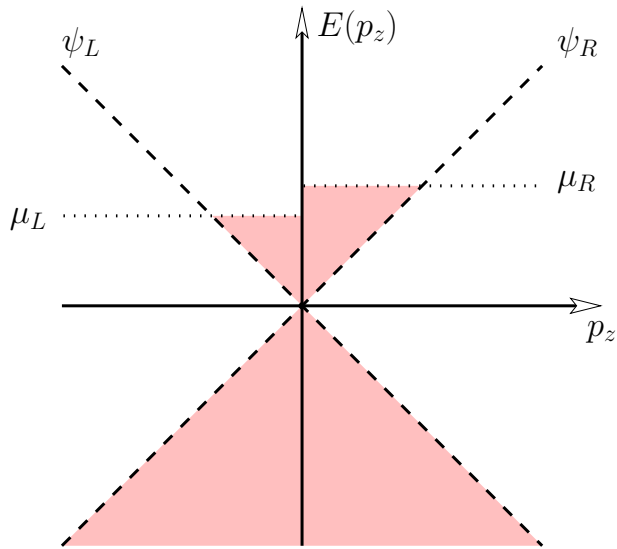}
  \caption{Lowest Landau level of the massless fermions filled up until the energy $\mu_L$ for $p_z <0$ and up to the energy $\mu_R$ for $p_z >0$.}
  \label{fig:LLL_CME}
\end{figure}

Now, let us introduce finite temperature and chemical potentials $\mu_L, \mu_R$
and fill both left- and right-chiral branches of the lowest Landau level according to the Fermi-Dirac distribution (see  Fig.~\ref{fig:LLL_CME}):
\begin{eqnarray}
  \label{eq:11}
&&  f_L(p) = \frac1{\exp\left(\frac{c|p_z| - \mu_L}{k_BT}\right) + 1}, \quad p_z < 0 \ ,
 \nonumber\\
&&   f_R(p) = \frac1{\exp\left(\frac{c|p_z| - \mu_R}{k_BT}\right) + 1}, \quad p_z > 0 \ ,
\end{eqnarray}
where $T$ is the temperature.

Determining the quantum mechanical expectation value of the electric current
$e \langle \psi| \gamma_z |\psi\rangle $,
and weighting it with the Fermi-Dirac distributions~\eqref{eq:12}, we find
\begin{eqnarray}
  \label{eq:12}
   J_{\rm CME}
  &=& \frac{e B}{(2\pi\hbar)^2} \Big[\int\limits^\infty_0 \frac{dp_z}{2\pi\hbar}\, \psi_{p_z}^\dagger \gamma_z \psi_{p_z} f_R(p_z)
\nonumber\\
 &&+ \int\limits_{-\infty}^0 \frac{dp_z}{2\pi\hbar}\, \psi_{p_z}^\dagger \gamma_z \psi_{p_z} f_L(p_z)\Big] \ ,
\end{eqnarray}
where $\psi_{p_z}$ are the eigenfunctions of the Dirac equation on
the lowest Landau level.
Computing Equation~\eqref{eq:12}, we find that an additional electric
current given by Equation~\eqref{eq:13} appears.

\section{Ideal MHD, Lorentz force and Galilean invariance}
\label{sec:lorentz-force}

In this Section we will remind the reader of the derivation of the MHD equations suitable for both relativistic and nonrelativistic case.
Most of this material can be found in different textbooks \citep{Landau-vol6,Melrose-QuantumPlasmadynamicsMagnetized,Weinberg-gravitation}.
We repeat the derivation to demonstrate the origin and structure of the Lorentz force in the case in which the electric current does not have a simple form of the Ohmic current given by Equation~\eqref{JJ}.
We limit ourselves to the case of the \emph{ideal} MHD, which is sufficient for our purposes.

The equation of ideal (nondissipative) MHD can be derived based on the energy--momentum conservation arguments (see \cite[Section~133-134] {Landau-vol6}, \cite[Section~1.4]{Melrose-QuantumPlasmadynamicsMagnetized}, \cite[\S 2.10]{Weinberg-gravitation}.

The total energy--momentum tensor $T^{\mu\nu}_\tot$ is the sum of
the  electromagnetic and matter parts: $  T^{\mu\nu}_\tot  \equiv
T^{\mu\nu}_{\mathrm{EM}}  + T^{\mu\nu}_\mat$, where the
energy--momentum tensor of the ideal fluid is given by
\begin{equation}
  \label{eq:44}
  T^{\mu\nu}_\mat =
    (\rho +p/c^2)
  u^\mu u^\nu + p \eta^{\mu\nu} ,
\end{equation}
where
$\rho c^2$ is \emph{energy density},
$p$ is pressure, $\eta_{\mu\nu}$ is the Minkowski metric (whose signature we choose to be $(-,+,+,+)$ and $u^\mu$
is the 4-velocity:
\begin{equation}
  \label{eq:46}
  u^\mu \equiv \left(\frac1{\sqrt{1-\UU^2/c^2}},\frac{\UU}{c \sqrt{1-\UU^2/c^2}}\right) = \gamma \,(1, \UU) ,
\end{equation}
where the $\gamma$-factor is defined via
\begin{equation}
\label{eq:50}
\gamma = \frac1{\sqrt{1-\UU^2/c^2}} .
\end{equation}
The electromagnetic stress-energy tensor $ T^{\mu\nu}_{\mathrm{EM}}$ is
\begin{equation}
  \label{eq:45}
   T^{\mu\nu}_{\mathrm{EM}} = F_{\mu\lambda}F_{\nu}^{\ \lambda} - \frac14 \eta_{\mu\nu} F_{\lambda\rho}F^{\lambda\rho} .
\end{equation}

The stress-energy tensor $T^{\mu\nu}_{\mathrm{EM}}$ is evolving as a
consequence of the Maxwell equations~\eqref{eq:maxwell}
\begin{equation}
  \label{eq:20}
  \frac{\partial  T^{\mu\nu}_\mathrm{EM}}{\partial  x^\nu}=-F^{\mu\nu}  J_{\tot,\,\nu} \, ,
\end{equation}
where the total current 4-vector $J^\nu_\tot \equiv (c\rho_\tot,\JJ_\tot)$.
We stress that the rhs of Equation~\eqref{eq:20} necessarily contains
the total current, which sources the Maxwell equations, independently
of its origin.

Taking the divergence of a spatial part ($\mu = i= 1,2,3$) of the matter-energy momentum
tensor~\eqref{eq:44}, one finds the relativistic generalization of the Euler equation:
\begin{small}
  \begin{equation}
    \label{eq:19}
    \pfrac{T^{i\nu}_\mat}{x^\nu} = \gamma^2
    (\rho +p/c^2)
    \left[\pfrac \UU t
    + \UU\cdot\nab \UU\right]_i + \left(\nab_i p + \frac{\UU_i}{c^2} \pfrac pt\right) .
  \end{equation}
\end{small}

The conservation of the $T^{\mu\nu}_\tot$ means that divergence~\eqref{eq:19} is equal to $F^{i\nu} J_{\tot,\,\nu}$, which coincides with the Lorentz force, and
  \begin{eqnarray}
  \label{eq:49}
    && \gamma^2 (\rho +p/c^2)\left[\pfrac \UU t + \UU\cdot\nab \UU\right]
    = -\nab p - \frac{\UU}{c^2} \pfrac pt
  \nonumber\\
    && \quad + \frac 1c\JJ_\tot\times\BB + \varrho_\tot \EE ,
  \end{eqnarray}
where the last term is zero if the plasma is electrically neutral (i.e.\ $\varrho_\tot = 0$).
This consideration carries an important assumption: there is no extra energy, associated with $\Theta$ field.

\subsection{Galilean invariance}
\label{sec:galilean-invariance}

Let us check that Equations~\eqref{eq:maxwell}--\eqref{eq:2} are Galilean invariant.
The Galilean transformation acts as follows:
  \begin{equation}
  \label{eq:42}
  \begin{aligned}
    \UU & \to \UU + \VV ,\\
    \BB & \to \BB+ \frac\VV c \times \EE , \\
    \EE&\to \EE - \frac\VV c \times \BB , \\
    \partial_t &\to \partial_t + \VV \cdot \nab ,\\
    \nab &\to \nab .\\
  \end{aligned}
\end{equation}

Expression~\eqref{eq:59} implies that the Ohmic current is given by Equation~\eqref{JJ}
\emph{in a comoving frame}, if we consider $\BB,\EE$ and $\UU$ in that particular frame.
It follows from Equation~\eqref{eq:59}
that there is a $\sigma (\EE \cdot \UU)$ contribution to the electric charge density $\varrho_\tot$
(again understanding $\EE$ and $\UU$ in a comoving frame).

The curl of $\BB$ in the \emph{comoving} frame is given by
\begin{eqnarray}
  \label{eq:58}
&& \nab\times \left(\BB+ \frac\VV c \times \EE\right) = \nab\times \BB +
  {1 \over c} \Big[\underbrace{(\nab\cdot \EE)}_{=4\pi\varrho_{\rm tot}} \VV
\nonumber\\
 && \quad - (\VV\cdot \nab) \EE \Big] .
 \end{eqnarray}
Therefore, the Galilean transformation for $\nab\times \BB$ is:
\begin{equation}
  \label{eq:57}
\nab\times \BB \to \nab\times \BB + \frac{4\pi}c \varrho_\tot\VV
- \frac{1}c (\VV\cdot \nab) \EE .
\end{equation}
The last term in Equation~\eqref{eq:57}, $-(1/c) (\VV\cdot \nab) \EE$, is a part of the displacement current in Equation~\eqref{rotB}, which is transformed as $\partial_t \EE \to (\partial_t + \VV \cdot \nab) \EE$.
The second term on the rhs of Equation~\eqref{eq:58} indicates that any
current should transform as
\begin{equation}
\label{eq:57a}
\JJ \to \JJ - \varrho_\tot \VV .
\end{equation}

Consider ordinary MHD equations, but keeping $\varrho_\tot \neq 0$.
Then we have
\begin{equation}
  \label{eq:77}
  \nab\times \BB = \frac{4\pi}c \Bigl[\sigma\left(\EE + \frac{\UU\times \BB}c\right)
   - \varrho_\tot \UU\Bigr] ,
\end{equation}
and
\begin{equation}
  \label{eq:74}
  \nab\cdot \EE = 4\pi\varrho_\tot .
\end{equation}
Expressing $\EE$ from Equation~\eqref{eq:77} we find
\begin{equation}
  \label{eq:78}
  \EE = -\frac{\UU\times \BB}c + \sigma^{-1} \left(\frac{c}{4\pi} \nab\times\BB
  +\varrho_\tot \UU\right) .
\end{equation}
This expression has correct transformation properties if one takes into account
Equation~\eqref{eq:57}.

\section{Procedure of the derivation of the mean electromotive force}
\label{sec:mean-emf}

We consider an incompressible turbulence with a zero mean kinetic helicity.
In order to derive equations for the nonlinear
coefficients defining the mean electromotive force, we will use a
mean-field approach.
Below we consider several methods
for the derivation of the equation for the nonlinear mean electromotive force.

\subsection{Quasi-linear approach}
\label{sec:quasi-linear approach}

First, we consider a weakly nonlinear case in which nonlinear terms in
Equations~(\ref{MB1})--(\ref{MB3}) are much smaller than viscous and diffusion terms.
This allows us to use the quasi-linear approach.
Using this approach, we neglect the nonlinear term in Equation~(\ref{MB2})
and keep the diffusion term.
We also use a multiscale approach, which allows us to separate
small-scale effects (fluctuations) from large-scale effects (mean fields).
In particular, let us calculate the function $\overline{b_i(t, {\bm x}) \, u_j(t, {\bm x})}$:
\begin{eqnarray}
&& \overline{b_i(t, {\bm x}) \, u_j(t, {\bm x})} =  \lim_{t_1 \to t_2, {\bm x} \to {\bm y}}
\overline{b_i(t_1, {\bm x}) \, u_j(t_2, {\bm  y})}
\nonumber\\
&& = \lim_{\tilde \tau \to 0, {\bm r} \to 0} \int  \,d {\bm k} \, \int \,d\omega \, \,
\overline{b_i(\omega, {\bm k}) \, u_j(-\omega, -{\bm k})}
\nonumber\\
&& \times
\exp[i {\bm k} {\bm \cdot} {\bm r} + i\omega \, \tilde \tau]
= \int \,d {\bm k} \int \,d\omega \, \, \overline{b_i(\omega, {\bm k}) \, u_j(-\omega, -{\bm k})} ,
\nonumber\\
\label{J11}
\end{eqnarray}
where we use new variables,
${\bm r} = {\bm x} - {\bm y}$, $\, \tilde \tau = t_1 - t_2$, $\, {\bm k} = ({\bm k}_1
- {\bm k}_2) / 2$, $\omega = (\omega_1 - \omega_2) / 2$, which correspond to small-scale
variables, and  ${\bm R} = ({\bm x} +  {\bm y}) / 2$, $t = (t_1 + t_2) / 2$, ${\bm K} =
{\bm k}_1 + {\bm k}_2$, $\Omega = \omega_1 + \omega_2$ which correspond to the large-scale
variables.
For inhomogeneous turbulence the correlation functions $\langle b_i \, u_j\rangle$,
$\langle u_i \, u_j\rangle$, etc.,
depend on the large-scale variable ${\bm R}$.
We assume also here that there exists a separation of scales,
i.e., the maximum scale of random motions $\ell_0$ is much
smaller than the characteristic scales of inhomogeneities of the
mean fields.

Equation~(\ref{MB2}) written in Fourier space yields
\begin{eqnarray}
\hat D_{in} \, b_n(\omega, {\bm k}) &=& G_\eta \Big[i({\bm k} {\bf \cdot} \meanBB) \delta_{im}
- \meanB_{i,m}\Big] u_m(\omega, {\bm k})
\nonumber\\
&& + G_\eta \, \eta \, \mu' \, \nab \times \meanBB,
\label{J12}
\end{eqnarray}
where $\hat D_{ij} = \delta_{ij} - \tilde \phi \, \varepsilon_{ijm} k_m$,
$\, \delta_{ij}$ is the Kronecker unit tensor,
$\, \tilde \phi = - i k \, G_\eta \, \meanv_\mu$, $\meanB_{i,j} = \nabla_j \meanB_{i}$,
and $G_\eta(k, \omega) = (\eta k^2 + i \omega)^{-1}$.
We consider here the case of a uniform chiral chemical potential, so
we neglected terms $\sim O(\nabla \meanmu)$ in Equation~(\ref{J12}),
where $\bec{\nabla} = \partial / \partial {\bm R} $.
Using Equation~(\ref{J12}) we derive an equation for the cross-helicity tensor
$g_{ij}=\overline{b_i(\omega, {\bm k}) u_j(-\omega, -{\bm k})}$:
\begin{eqnarray}
g_{ij} = G_\eta \, \hat D_{il}^{-1} \, \Big[i({\bm k} {\bf \cdot} \meanBB) \delta_{lm}
- \meanB_{l,m}\Big] f_{mj} ,
\label{J14}
\end{eqnarray}
where $f_{ij}=\overline{u_i(\omega, {\bm k}) u_j(-\omega, -{\bm k})}$ and the operator
\begin{eqnarray}
\hat D_{ij}^{-1} = [\delta_{ij} +
\tilde \phi \, \varepsilon_{ijm} \, (k_m/k) + \tilde \phi^2 \, k_{ij}]/(1 + \tilde \phi^2)
\label{J15}
\end{eqnarray}
is the inverse of $\hat D_{ij}$, i.e.,  $\hat D_{in}^{-1} \, \hat D_{nj} = \delta_{ij}$.
In Equation~(\ref{J14}) we neglected terms $\sim O[\meanB^2; (\nabla \meanBB)^2;
\nabla^2 \meanBB]$.
The mean electromotive force is defined as $\meanEMF\equiv \overline{{\bm u} \times {\bb}}= \varepsilon_{mji} \int g_{ij}(\omega,{\bm k}) \,d {\bm k} \, d\omega$.
The contribution to the mean electromotive force $\meanEMF^\mu$ caused by the chiral effect is obtained using Equation~(\ref{J14}):
\begin{eqnarray}
\overline{\cal E}^\mu_{m} = \left(\int a_{ij}^\mu (\omega,{\bm k}) \,d\omega \,d {\bm k}\right) \,
\meanB_j ,
\label{J16}
\end{eqnarray}
where
\begin{eqnarray}
a_{ij}^\mu(\omega,{\bm k}) = - \meanv_\mu \, G_\eta^2 k_i k_j f^{(0)}_{mm}(\omega,{\bm k})
+ O(\meanv_\mu^2) ,
\label{J17}
\end{eqnarray}
The correlation functions with the superscript $(0)$ correspond to the background turbulence
with a zero mean magnetic field.
To integrate in $\omega$ and ${\bm k}$ space, we use the following model for the background isotropic, homogeneous and nonhelical turbulence:
\begin{eqnarray}
f_{ij}^{(0)}(\omega, {\bm k}) &=& {\tilde E(k) \, \tilde \Phi(\omega)
\over 8 \pi \, k^2} \Big[\delta_{ij}
- {k_i \, k_j \over k^2} \Big] \overline{{\bm u}^2},
\label{J20}
\end{eqnarray}
where the energy spectrum function $\tilde E(k) =  (q-1) k_0^{-1} (k /k_{0})^{-q}$
with the exponent $1<q<3$ and for $k_0 < k < k_d$.
Here $k_{0} = 1 / \ell_{0}$, $k_d$ is the wavenumber based on the dissipation scale,
and $k_0 \ll k_d$.
We consider  the frequency function $\tilde \Phi(\omega)$
in the form of the Lorentz profile:
$\tilde \Phi(\omega)=1 / [\pi \, \tau_c \,(\omega^2 + \tau_c^{-2})]$,
where $\tau_c$ is the characteristic correlation time of random velocity field.
This model for the frequency function
corresponds to the correlation function
\begin{eqnarray}
\langle u_i(t) u_j(t+\tau) \rangle \propto \exp (-\tau /\tau_c).
\end{eqnarray}
After integration in $\omega$ and ${\bm k}$ space in Equations~(\ref{J16}),
we obtain the contribution to the mean
electromotive force $\meanEMF^\mu$ caused by the CME
$\meanEMF^\mu = \alpha_\mu  \meanBB$,
where the $\alpha_\mu$ effect is determined by Equation~(\ref{J19}).

\subsection{The $\tau$ approach}
\label{sec:tau approach}

In this section we consider the case of large hydrodynamic and magnetic Reynolds numbers.
This implies that the nonlinear terms in Equations~(\ref{MB1})--(\ref{MB3})
are much larger than viscous and diffusion terms.
To exclude the pressure term from
the equation of motion~(\ref{MB1}) we calculate
$\bec{\nabla} {\bf \times} (\bec{\nabla} {\bf \times} {\bm u})$.
Using Equations~(\ref{MB1})--(\ref{MB3}) written in Fourier space, we
derive equations for the correlation functions of the velocity
field $f_{ij}=\overline{u_i u_j}$, the magnetic field
$h_{ij}=\overline{b_i b_j}$, the cross-helicity
$g_{ij}=\overline{b_i u_j}$, the flux of chemical potential
$s_{j}=\overline{\mu' u_j}$ and
the correlation $q_{i}=\overline{\mu' b_i}$:
\begin{eqnarray}
{\partial f_{ij}({\bm k}) \over \partial t} &=& i({\bm k} {\bf
\cdot} \meanBB) \Phi_{ij} + I^f_{ij}
+ F_{ij} + f_{ij}^N ,
\label{MB5} \\
{\partial h_{ij}({\bm k}) \over \partial t} &=& - i({\bm k}{\bf
\cdot} \meanBB) \Phi_{ij} + I^h_{ij} + h_{ij}^N ,
\label{MB6} \\
{\partial g_{ij}({\bm k }) \over \partial t} &=& i({\bm k} {\bf
\cdot} \meanBB) [f_{ij}({\bm k}) - h_{ij}({\bm k}) -
h_{ij}^{(H)}] + I^g_{ij}
\nonumber\\
&&+ i \eta \varepsilon_{inm} k_n \left[s_j({\bf k})\meanB_m +
g_{mj}({\bm k})\meanmu\right] + g_{ij}^N \;,
\nonumber\\
\label{MB7}\\
{\partial s_{j}({\bm k}) \over \partial t} &=&
-i ({\bm k}{\bf \cdot} \meanBB) q_{j} + I^s_{j} + s_{j}^N + O(\eta),
\label{MB8}\\
{\partial q_{i}({\bm k}) \over \partial t} &=&
-i ({\bm k}{\bf \cdot} \meanBB) s_{i} + I^q_{i} + q_{i}^N + O(\eta),
\label{MB88}
\end{eqnarray}
where hereafter we omitted arguments $t$ and ${\bm R}$ in the
correlation functions, the mean fluid density is
included in the definition of the magnetic field, so that
the magnetic field is measured in units of the Alfv\'{e}n speed,
and we neglected terms $\sim O(\nabla^2)$.
Here $\Phi_{ij}({\bm k}) = g_{ij}({\bm k}) - g_{ji}(-{\bm
k}) ,$ $ \, F_{ij}({\bm k}) = \overline{\tilde F_i ({\bf k})
u_j(-{\bm k})} + \overline{u_i({\bm k}) \tilde F_j(-{\bm k
})} ,$ and ${\bm \tilde F} ({\bm k}) = {\bm k} {\bf \times}
({\bm k} {\bf \times} {\bm F}({\bm k})) / k^2$.
The source terms,
$I_{ij}^f, \, I_{ij}^h, \, I_{ij}^g, ...$, which contain the
large-scale spatial derivatives of the mean fields, are
given by
\begin{eqnarray}
&& I_{ij}^f = {1 \over 2}(\meanBB {\bf \cdot} \bec{\nabla})
\Phi_{ij}^{(P)} + [g_{qj}({\bm k}) (2 P_{in}(k) - \delta_{in})
\nonumber\\
&& \; + g_{qi}(-{\bm k}) (2 P_{jn}(k) - \delta_{jn})] \meanB_{n,q} -
\meanB_{n,q} k_{n} \Phi_{ijq}^{(P)},
\label{MB9}\\
&& I_{ij}^h = {1 \over 2}(\meanBB {\bf \cdot} \bec{\nabla})
\Phi_{ij}^{(P)} - [g_{iq}({\bm k}) \delta_{jn} + g_{jq}(-{\bm k})
\delta_{in}] \meanB_{n,q}
\nonumber\\
&& \; - \meanB_{n,q} k_{n} \Phi_{ijq}^{(P)} ,
\label{MB10}\\
&& I_{ij}^g = {1 \over 2} (\meanBB {\bf \cdot} \bec{\nabla})
(f_{ij} + h_{ij}) + h_{iq} (2 P_{jn}(k) - \delta_{jn})
\meanB_{n,q}
\nonumber\\
&& \; - f_{nj} \meanB_{i,n} - \meanB_{n,q} k_{n}(f_{ijq} + h_{ijq}) + \eta \varepsilon_{inm} g_{mj}({\bm k}) \nabla_n \meanmu,
\nonumber\\
\label{MB11}\\
&& I_{j}^s = - f_{nj} \nabla_n \meanmu + O(\eta),
\label{MB12}\\
&& I_{i}^q = - s_{n} \meanB_{i,n} - g_{in}(-{\bm k}) \nabla_n \meanmu
+ O(\eta),
\label{MB80}
\end{eqnarray}
where $\Phi_{ij}^{(P)}({\bm k}) = g_{ij}({\bm k}) + g_{ji}(-{\bm k})$,
$P_{ij}(k)=\delta_{ij} - k_i \, k_j / k^2$,
the terms $f_{ij}^{N} , \, h_{ij}^{N}, \, g_{ij}^{N}, ... $ are determined
by the third moments appearing as a result of the nonlinear terms
(these terms also include the dissipative viscous and diffusion terms),
$f_{ijq} = (1/2) \partial f_{ij} / \partial k_{q} ,$ and similarly
for $h_{ijq}$ and $\Phi_{ijq}^{(P)}$.
A stirring force in the
Navier-Stokes turbulence is an external parameter, which determines
the background turbulence.
We have taken into account that in Equation (\ref{MB7}) the terms with
symmetric tensors with respect to the indexes ``i'' and ``j'' do not
contribute to the mean electromotive force because ${\cal E}_{m} =
(1/2)\varepsilon_{mji} \, \Phi_{ij} $.
In Equations (\ref{MB9})--(\ref{MB80}) we
have neglected the second and higher derivatives over $ {\bf R}$.

In Equations~(\ref{MB6}) and (\ref{MB7}) we split the tensor
for magnetic fluctuations into nonhelical, $h_{ij},$ and helical,
$h_{ij}^{(H)},$ parts.
The helical part of the tensor $h_{ij}^{(H)}$ for magnetic
fluctuations depends on the magnetic helicity, and it is
determined by the dynamic equation that follows from the magnetic
helicity conservation arguments
\citep{KR82,KRR95,GD94,KR99,KMRS2000,BF00,BB02,BS05}.
The characteristic time of evolution of the nonhelical part of the magnetic tensor
$h_{ij}$ is of the order of the turbulent correlation time $
\tau_{0} = \ell_{0} / u_{0}$, while the relaxation time of the
helical part of the magnetic tensor $h_{ij}^{(H)}$ is of the
order of $ \tau_{0} \, \Rm $ \citep{KRR95,KR99},
where $\Rm = \ell_0 u_{0} / \eta \gg 1$ is the magnetic
Reynolds number and $u_{0}$ is the characteristic turbulent velocity
in the maximum scale $\ell_{0}$ of turbulent motions.

The equations for the second moment include the first-order spatial
differential operators applied to the third-order
moments.
A problem arises regarding how to close the system, i.e.,
how to express the third-order terms $\hat{\cal N}
F^{(III)}$ through the lower moments $F^{(II)}$
\citep{O70,MY75,Mc90}.
We use the spectral $\tau$ approximation, which postulates that the deviations of the third-moment terms, $\hat{\cal N} F^{(III)}({\bm k})$, from the contributions to these terms by the background turbulence, $\hat{\cal N} F^{(III,0)}({\bm k})$, can be expressed through similar deviations of the second moments, $F^{(II)}({\bm k}) - F^{(II,0)}({\bm k})$ \citep{O70,PFL76,KRR90,KMR96}:
\begin{eqnarray}
&& \hat{\cal N} F^{(III)}({\bm k}) - \hat{\cal N} F^{(III,0)}({\bm
k})
\nonumber\\
&& \; \;= - {1 \over \tau_r(k)} \, \Big[F^{(II)}({\bm k}) - F^{(II,0)}({\bm k})\Big] ,
\label{MB13}
\end{eqnarray}
where $\tau_r(k)$ is the scale-dependent relaxation time, which can be identified
with the correlation time $\tau(k)$ of the turbulent velocity field for large fluid
and magnetic Reynolds numbers.
The functions with the superscript $(0)$ correspond to the background turbulence
with a zero mean magnetic field.
Validation of the $\tau$ approximation for different situations has been performed in numerous numerical simulations and analytical studies \citep{BS05,RK07,RKKB11}.

In this study we consider an intermediate nonlinearity that
implies that the mean magnetic field is not strong enough in order
to affect the correlation time of the turbulent velocity field.
The theory for a very strong mean magnetic field can be corrected
after taking into account the dependence of the correlation time of
the turbulent velocity field on the mean magnetic field.

\subsubsection{Solution of second-moment equations}

We start with Equations~(\ref{MB5})--(\ref{MB88}) written for nonhelical parts of the
tensors.
We subtract Equations (\ref{MB5})--(\ref{MB88}) written for
background turbulence (for $\meanBB=0)$ from those for
$\meanBB \not=0$.
Then we use the $\tau$ approximation.
Next, we assume that $\eta k^2 \ll \tau^{-1}(k)$ and $\nu k^2 \ll
\tau^{-1}(k)$ for the inertial range of turbulent flow, and we also
assume that the characteristic time of variation of the mean
magnetic field $\meanBB$ is substantially larger than the
correlation time $\tau(k)$ for all turbulence scales.
Thus, we
arrive to the following steady-state solution of the obtained
equations:
\begin{eqnarray}
f_{ij}({\bm k}) &=& f_{ij}^{(0)}({\bm k}) + i \tau
({\bm k} {\bf \cdot} \meanBB) \Phi_{ij}({\bm k}) + \tau I_{ij}^f,
\label{MB14}\\
h_{ij}({\bm k}) &=& h_{ij}^{(0)}({\bm k}) - i \tau
({\bm k} {\bf \cdot} \meanBB) \Phi_{ij}({\bm k}) + \tau I_{ij}^h,
\label{MB15}\\
g_{ij}({\bm k}) &=& i \tau ({\bm k} {\bf \cdot}
\meanBB) \left[f_{ij}({\bm k}) - h_{ij}({\bm k})\right]
+ \tau \Big\{I_{ij}^g
\nonumber\\
&& + i \eta \varepsilon_{inm} k_n \left[s_j({\bf k})\meanB_m +
g_{mj}({\bm k})\meanmu\right]\Big\},
\label{MB16}\\
s_{j}({\bm k}) &=& -i \tau ({\bm k}{\bf \cdot} \meanBB) q_{j} + \tau I_{j}^s + O(\eta),
\label{MB17}\\
q_{j}({\bm k}) &=& -i \tau ({\bm k}{\bf \cdot} \meanBB) s_{j} + \tau I^q_{j} ,
\label{MB81}
\end{eqnarray}
where we have taken into account that $g_{ij}^{(0)}({\bm k}) = 0$.
Equations~(\ref{MB14})--(\ref{MB81}) yield
\begin{eqnarray}
&& \Phi_{ij}= \hat \Phi_{ij} + \hat \Phi_{ij}^\mu+ \Phi_{ij}^{(I)} ,
\label{MB18}\\
&& \hat \Phi_{ij}({\bm k}) = {2i \tau ({\bm k} {\bf \cdot}
\meanBB) \over 1 + 2 \Psi} [f_{ij}^{(0)}({\bm k}) -
h_{ij}^{(0)}({\bm k})] = 2 \hat g_{ij}({\bm k}),
\nonumber\\
\label{MB19}\\
&& \hat \Phi_{ij}^\mu = i \eta \tau \left[\varepsilon_{inm} \hat g_{mj}({\bm k})
+ \varepsilon_{jnm} \hat g_{mi}(-{\bm k})\right]\, k_n \, \meanmu ,
\label{MA19}
\end{eqnarray}
\begin{eqnarray}
&& \Phi_{ij}^{(I)}({\bf k}) = {\tau \over 1 + 2 \Psi} \Big[
\tilde I^g_{ij}({\bf k}) - \tilde I^g_{ji}(-{\bf k})
\nonumber\\
&& \quad + 2 i \tau ({\bf k} {\bf \cdot} \bar{\bf B})\left(I^f_{ij} - I^h_{ij}\right) \Big],
\label{MB20}
\end{eqnarray}
where $ \Psi({\bm k}) = 2 (\tau \, {\bm k} {\bf \cdot} \meanBB)^2$,
\begin{eqnarray}
\tilde I^g_{ij}({\bm k}) = I_{ij}^g + i \eta \varepsilon_{inm} k_n s_j^{(I)} \meanB_m,
\label{MB21}
\end{eqnarray}
and $\hat f_{ij}, \hat h_{ij}, ...$ are the solutions of Equations~(\ref{MB14})--(\ref{MB81})
without the source terms $I_{ij}^f, \, I_{ij}^h, \, I_{ij}^g, ...$ caused by the gradients
of the mean fields,
and $f_{ij}^{(I)}, h_{ij}^{(I)}, g_{ij}^{(I)}, ...$ are the solutions of these equations
caused by the source terms.
For example, the function $s_j^{(I)}$ entering in Equation~(\ref{MB21})
is given by
\begin{eqnarray}
s_j^{(I)}({\bf k})= {\tau I_{j}^s \over 1 + \Psi/2} = -{\tau \hat f_{nj} \nabla_n \meanmu \over 1 + \Psi/2} + O(\eta).
\label{MB82}
\end{eqnarray}
In derivation of Equations~(\ref{MB19})--(\ref{MB21}) we have taken into account
the following arguments.
Since the solution for the correlation function
$\hat s_{j}({\bm k})$ is proportional to $\eta$,
and since we should neglect effects that are quadratic in $\eta$,
the correlation function $\hat s_{j}({\bm k})$ does not contribute to
$\hat \Phi_{ij}({\bm k})$.
Using Equations~(\ref{MB14})--(\ref{MB16}) and~(\ref{MB19}) we obtain
\begin{eqnarray}
\hat f_{ij}({\bm k}) &\approx& {1 \over 1 + 2 \Psi} [(1 +
\Psi) f_{ij}^{(0)}({\bm k}) + \Psi h_{ij}^{(0)}({\bm k})] ,
\label{MB25} \\
\hat h_{ij}({\bm k}) &\approx& {1 \over 1 + 2 \Psi} [\Psi
f_{ij}^{(0)}({\bm k}) + (1 + \Psi) h_{ij}^{(0)}({\bm k})] .
\label{MB26}
\end{eqnarray}

Using the derived equations for the second moments $f_{ij}, \,
h_{ij}, g_{ij}, ...$ we calculate the mean electromotive force
${\cal E}_{l} = \int \tilde {\cal E}_{l}({\bm k}) \,d {\bm k} ,$
where $ \tilde{\cal E}_{l}({\bm k}) = (1/2)\varepsilon_{lji}
\Phi_{ij}({\bm k})$.
The total mean electromotive force is $\bec{\cal E}=\meanEMF+\meanEMF^\mu$,
where $\meanEMF$ are the contributions to the mean electromotive force
without the CME, while $\meanEMF^\mu$ are
the contributions to the mean electromotive force caused by the CME.
The contribution $\meanEMF^\mu$ is determined using Equations~(\ref{MB20})--(\ref{MB25}):
\begin{eqnarray}
&& \overline{\cal E}^\mu_{l} = {1\over 2}\varepsilon_{lji} \int \hat \Phi_{ij}^\mu  ({\bm k}) \,d {\bm k}
\nonumber\\
&& \; \;
- i \eta \,\varepsilon_{lji} \, \varepsilon_{inm} \, \meanB_m \int {\tau(k) k_n s_j^{(I)}({\bm k}) \over 1 + 2 \Psi} \,d {\bm k} .
\label{MB30}
\end{eqnarray}
The first term in Equations~(\ref{MB30}) determines the contribution to $\meanEMF^\mu$ caused by homogeneous turbulence with uniform  chiral chemical potential, while  the second term in Equation~(\ref{MB30}) determines the contribution to $\meanEMF^\mu$ caused by inhomogeneous turbulence with nonuniform chiral chemical potential.

\subsubsection{Mean electromotive force in homogeneous nonhelical turbulence}

We use the following model for the background
isotropic, homogeneous, and nonhelical turbulence:
\begin{eqnarray}
f_{ij}^{(0)}({\bm k}) &=& {\tilde E(k)
\over 8 \pi \, k^2} \Big[\delta_{ij}
- {k_i \, k_j \over k^2} \Big] \overline{{\bm u}^2},
\label{C10}
\end{eqnarray}
where the turbulent time $\tau(k) = 2 \tau_{0} \tilde \tau(k)$ with $\, \tilde\tau(k) = (k /k_{0})^{-2/3}$, the energy spectrum function $\tilde E(k) = - (d \tilde \tau(k) / dk)
= (2/3) k_0^{-1} (k /k_{0})^{-5/3}$ corresponds to the Kolmogorov turbulence,
$k_{0} = 1 / \ell_{0}$  and $\tau_{0} = \ell_{0} / u_{0}$.
We also assumed for simplicity that $h_{ij}^{(0)}({\bm k})=0$
(i.e., no small-scale dynamo).

After the integration in ${\bm k}$ space, we obtain the contributions to the mean electromotive
force caused by the uniform chiral chemical potential: $\meanEMF^\mu=\alpha_\mu \meanB$,
where the $\alpha_\mu$ effect is determined by Equation~(\ref{MB130A}) for
magnetic Prandtl numbers $\Pm \geq 1$
and by Equation~(\ref{MB130B}) for $\Pm < 1$.

\subsubsection{Mean electromotive force in inhomogeneous nonhelical turbulence}

Now we use the model for the background isotropic, inhomogeneous,
and nonhelical turbulence:
\begin{eqnarray}
f_{ij}^{(0)}({\bm k}) &=& {\tilde E(k)
\over 8 \pi \, k^2} \Big[\delta_{ij}
- {k_i \, k_j \over k^2} + {i \over 2 k^2} \, \big(k_i \nabla_j - k_j \nabla_i\big)
\Big] \overline{{\bm u}^2} .
\nonumber\\
\label{C11}
\end{eqnarray}

After the integration in ${\bm k}$ space, we obtain
the contributions to the mean electromotive force caused by the nonuniform chiral chemical
potential:
\begin{eqnarray}
\meanEMF^\mu= {\eta_{_{T}} \eta \, \tau_0 \over 18} \, S(\beta)
\Big[\left(\nabla_i \meanmu \right)\, \meanBB {\bm \cdot} \nab
+ \meanB_i \left(\nabla_p \meanmu \right)\nabla_p\Big]\ln \overline{{\bm u}^2},
\nonumber\\
\label{MB100}
\end{eqnarray}
where
\begin{eqnarray}
S(\beta) &=& 28 A_{1}(\sqrt{2} \beta) - A_{1}(\beta/\sqrt{2}) -
{9 \over \pi} \bar  A_{1}(2 \beta^2),
\label{MB105}\\
A_{1}(x) &=& {6 \over 7} \biggl[{\arctan x \over x} \biggl(1 + {7 \over 9 x^{2}} \biggr) + {1 \over 27} - {7 \over 9x^{2}}
\nonumber\\
&& - {x^{2} \over 18} \left[1 - 2 x^{2} + 2 x^{4} \ln (1 + x^{-2})\right] \biggr]  ,
\label{MB105a}\\
 \bar A_{1}(x) &=& {2 \pi \over x} \biggl[(x + 1) {\arctan (\sqrt{x}) \over
\sqrt{x}} - 1 \biggr] .
\end{eqnarray}
For $x \ll 1 $ these functions are given by
\begin{eqnarray*}
A_{1}(x) &\sim& {2 \over 3}\biggl(1 - {3 \over 10}
x^{2} \biggr) , \; \bar A_{1}(x) \sim {4 \pi \over 3} \biggl(1 - {1 \over 5} x
\biggr) ,
\end{eqnarray*}
and for $x \gg 1 $ they are given by
\begin{eqnarray*}
A_{1}(x) &\sim& {3 \pi \over 7 x} - {3 \over 2x^2} + {\pi \over 3 x^3},
\; \bar A_{1}(x) \sim {\pi^{2} \over x^{1/2}} - {4 \pi \over x} + {\pi^{2} \over x^{3/2}}.
\end{eqnarray*}
Using these asymptotic formulae, we determine the contributions to the mean electromotive force caused by the nonuniform chiral chemical potential for weak and strong mean magnetic fields.
For the weak field, $\beta \ll 1$, the contributions to the mean electromotive force $\overline{\cal E}_i^\mu$ are given by
\begin{eqnarray}
&& \overline{\cal E}_i^\mu= {\eta_{_{T}} \eta \, \tau_0 \over 3} \, \left(1-{21 \over 20}\beta^2\right)
\Big[\left(\nabla_i \meanmu \right)\, \meanBB {\bm \cdot} \nab
\nonumber\\
&& \quad + \meanB_i \left(\nabla_p \meanmu \right)\nabla_p\Big]\ln \overline{{\bm u}^2},
\label{MB109}
\end{eqnarray}
and for strong field $\beta \gg 1$ they are given by
\begin{eqnarray}
&& \overline{\cal E}_i^\mu= {\eta_{_{T}} \eta \, \ell_0 \over 11 \meanB} \,
\Big[\left(\nabla_i \meanmu \right)\, \meanBB {\bm \cdot} \nab
+ \meanB_i \left(\nabla_p \meanmu \right)\nabla_p\Big]\ln \overline{{\bm u}^2} .
\nonumber\\
\label{MB112}
\end{eqnarray}
We remind here that the mean fluid density is included in the definition
of the magnetic field, so that the magnetic field is measured in units of the Alfv\'{e}n speed,
and the electromotive force $\overline{\cal E}_i$ is measured in the
units of the squared velocity, while $\meanmu$ is measured in the units of the inverse scale.

The total mean electromotive force can be written in the form
\begin{eqnarray}
\overline{\cal E}_i = a_{ij} \, \meanB_{j} + b_{ijk} \, \meanB_{j,k} \;,
\label{MB40}
\end{eqnarray}
where $\meanB_{j,i} = \nabla_i \meanB_{j}$ and we neglected terms
$\sim O(\nabla^2 \meanB_{k})$.
The general form of the mean electromotive force is given by Equation~(\ref{MB41}),
where the turbulent transport coefficients are related to
the tensors $a_{ij}$ and $b_{ijk}$:
\begin{eqnarray}
&& \alpha_{ij}(\bar{\bf B}) = {1 \over 2}(a_{ij} + a_{ji}) ,
\; V^{\rm eff}_k(\bar{\bf B}) = {1 \over 2} \varepsilon_{kji}
\, a_{ij} ,
\label{MB42} \\
&& \eta_{ij}(\bar{\bf B}) = {1 \over 4}(\varepsilon_{ikp} \,
b_{jkp} + \varepsilon_{jkp} \, b_{ikp}) ,
\label{MB43}\\
&& \delta_{i} = {1\over 4} (b_{jji} - b_{jij}),
\label{MMB43}\\
&&\kappa_{ijk}(\bar{\bf B}) = - {1 \over 2}(b_{ijk} + b_{ikj}) .
\label{MB44}
\end{eqnarray}
The separation of terms in Equations~(\ref{MB42}) and (\ref{MB43}) is not unique,
because a gradient term can always be added to the electromotive force.
Using Equations~(\ref{MB100}), (\ref{MB105}), (\ref{MB40}), and (\ref{MB42}),
we determine the functions $\alpha_{ij}^\mu(\meanBB)$ and ${\bm V}_{\rm eff}^\mu(\meanBB)$
caused by the nonuniform chiral chemical potential,
which are given by Equations~(\ref{MB150})--(\ref{MB159}).

\bibliographystyle{apj}%
\bibliography{paper-1-chiral-mhd}%

\end{document}